\documentclass[useAMS,usenatbib]{mn2e}
\usepackage{graphicx}
\usepackage{xcolor}
\usepackage{multirow}
\usepackage{wasysym}  
\usepackage{sidecap}
\usepackage{amssymb}

\title[Phase resolved spectroscopy and {\it Kepler} photometry of SDSS\,J190817.07+394036.4]{Phase resolved spectroscopy and {\it Kepler} photometry of the ultracompact AM\,CVn binary SDSS\,J190817.07+394036.4}
\author[T. Kupfer et al.]{T. Kupfer$^{1}$\thanks{E-mail:t.kupfer@astro.ru.nl}, P.~J. Groot$^{1,2}$, S. Bloemen$^{1}$, D. Levitan$^{2}$, D. Steeghs$^{3}$, T.~R. Marsh$^{3}$,  \newauthor{R.~G.~M. Rutten$^{4,5}$, G. Nelemans$^{1,6}$, T. A. Prince$^{2}$, F. F\"urst$^{2}$ and S. Geier$^{8}$}\\
$^{1}$Department of Astrophysics/IMAPP, Radboud University Nijmegen, P.O. Box 9010, 6500 GL Nijmegen, The Netherlands\\
$^{2}$Division of Physics, Mathematics, and Astronomy, California Institute of Technology, Pasadena, CA 91125, USA\\
$^{3}$Department of Physics, University of Warwick, Coventry CV4 7AL, UK\\
$^{4}$GRANTECAN, Center for Astrophysics in La Palma, 38712 Bre$\tilde{n}$a Baja, Spain\\
$^{5}$Gemini Observatory, Casilla 603, La Serena, Chile\\
$^{6}$Institute for Astronomy, KU Leuven, Celestijnenlaan 200D, 3001 Leuven, Belgium\\
$^{7}$Dr. Karl Remeis-Observatory \& ECAP, Friedrich-Alexander University Erlangen-Nuremberg, Sternwartstr. 7, 96049 Bamberg, Germany\\
$^{8}$European Southern Observatory, Karl-Schwarzschild-Str. 2, 85748 Garching, Germany}
\begin{document}

\date{Accepted --- Received ----; in original form ---}

\pagerange{\pageref{firstpage}--\pageref{lastpage}} \pubyear{2002}

\maketitle

\label{firstpage}

\begin{abstract}
{\it Kepler} satellite photometry and phase-resolved spectroscopy of the ultracompact AM\,CVn type binary SDSS\,J190817.07+394036.4 are presented. The average spectra reveal a variety of weak metal lines of different species, including silicon, sulphur and magnesium as well as many lines of nitrogen, beside the strong absorption lines of neutral helium. The phase-folded spectra and the Doppler tomograms reveal an S-wave in emission in the core of the He\,{\sc i} 4471\,\AA\,absorption line at a period of $P_{\rm orb}=1085.7\pm2.8$\,sec identifying this as the orbital period of the system. The Si\,{\sc ii}, Mg\,{\sc ii} and the core of some He\,{\sc i} lines show an S-wave in absorption with a phase offset of $170\pm15^\circ$ compared to the S-wave in emission. The N\,{\sc ii}, Si\,{\sc iii} and some helium lines do not show any phase variability at all. The spectroscopic orbital period is in excellent agreement with a period at $P_{\rm orb}=1085.108(9)$\,sec detected in the three year {\it Kepler} lightcurve. A Fourier analysis of the Q6 to Q17 short cadence data obtained by {\it Kepler} revealed a large number of frequencies above the noise level where the majority shows a large variability in frequency and amplitude. In an O-C analysis we measured a $\vert\dot{P}\vert\sim1.0\,$x$\,10^{-8}\,$s\,s$^{-1}$ for some of the strongest variations and set a limit for the orbital period to be $\vert\dot{P}\vert<10^{-10}$s\,s$^{-1}$. The shape of the phase folded lightcurve on the orbital period indicates the motion of the bright spot. Models of the system were constructed to see whether the phases of the radial velocity curves and the lightcurve variation can be combined to a coherent picture. However, from the measured phases neither the absorption nor the emission can be explained to originate in the bright spot.


\end{abstract}


\section{Introduction}
AM\,CVn systems are a small group of mass transferring ultracompact binaries with orbital periods between 5.4 and 65 minutes. AM\,CVn systems consist of a white dwarf (WD) primary and a WD or semi-degenerate helium star secondary (\citealt{nel01}, see \citet{sol10} for a recent review). They are predicted to be strong, low-frequency, Galactic gravitational wave sources (e.g. \citealt{nel04, roe07b, nis12}), the source population of the proposed “.Ia” supernovae \citep{bil07}, and as probes of the final stages of binary evolution. Spectroscopically these systems are characterized by a deficiency of hydrogen, indicating an advanced stage of binary evolution. In the preceding binary evolution two common envelope phases or a stable Roche Lobe overflow + one common envelope formed a detached WD binary system at a period of $\sim$hours. Gravitational wave radiation decreased the orbital separation until the low-mass secondary filled its Roche lobe and mass transfer set in at an orbital period between $3 - 10$\,minutes. Some fraction of these systems survived the ensuing direct impact phase to become AM\,CVn systems \citep{nel01,mar04} depending on their mass ratio and the efficiency of the angular momentum feedback. An accretion disc forms at an orbital period of $\sim$10~min and the mass transfer stream hits the disc at the so-called bright spot. The mass-transfer-rate drops as the orbit widens and the system ends up as a more massive WD with an extremely low-mass WD ($\sim0.01$~M$_{\odot}$) at orbital periods of $40$--$60$\,min. 

However, the number of known longer period systems (P$_{\rm orb}>20$\,min) has seen a surge in recent years due to large scale synoptic surveys such as SDSS (e.g. \citealt{and05,and08,roe05,roe09,car14a}), PTF \citep{lev11,lev13,lev14}, and most recently Gaia (Campbell et al. in prep.). The number of known systems at the short orbital period end (P$_{\rm orb}<20$\,min) is limited to five. The most recently discovered, supposedly short period system, is SDSS\,J190817.07+394036.4 (hereafter SDSS\,J1908).

\begin{table}
 \centering
 \caption{Summary of the observations of SDSS\,J1908}
  \begin{tabular}{lcrc}
  \hline
  Telescope/Date & & N$_{\rm exp}$ & Exp. time (s) \\
   \hline\hline
  \multicolumn{4}{l}{{\it Kepler} satellite} \\
  2010/06/24 -- 2013/05/11 & & 1.3\,M & 60 \\
     \noalign{\smallskip}
  \multicolumn{4}{l}{WHT+ISIS (R1200B/R1200R)} \\
  \smallskip
2011/07/03 -- 2011/07/07 & & 1875 & 60 \\
     \noalign{\smallskip}
      \multicolumn{4}{l}{GTC+OSIRIS (R2000B)} \\
  2011/09/15 -- 2011/09/17 & & 369 & 60 \\    
   \noalign{\smallskip}
    \multicolumn{4}{l}{Keck+ESI (Echellette mode)} \\
 2012/07/12  & & 203 & 60 \\  
    \noalign{\smallskip}
    \multicolumn{4}{l}{Keck+ESI (Echellette mode)} \\
 2014/06/01  & & 4 & 900 \\  
      \hline
\end{tabular}
\label{observ}
\end{table}

SDSS\,J1908 was observed in the Sloan Digital Sky Survey (SDSS) as a relatively bright ($g=16.08$\,mag) blue object. The system was labeled as a possible compact pulsator and included by the Kepler Astroseismic Science Consortium (KASC) for the survey phase at short cadence in the Kepler Space observatory \citep{gil10,oes11}. A first detailed study of the object based on short cadence {\it Kepler} data obtained during quarter 3.3 is presented in \citet{fon11}, hereafter F11, where it was concluded that SDSS\,J1908 is a high state AM\,CVn system, similar to the prototype system AM\,CVn itself (see e.g. \citealt{roe06a}). From the spectroscopic analysis F11 found that the system is seen at an inclination angle between 10$^{\circ}$ and 20$^{\circ}$. The estimated mass transfer rate lies in the range $3.5 - 8.5\times$10$^{-9}$M$_{\odot}$~yr$^{-1}$ and the distance to the system is in the range $250 - 330$\,pc. The luminosity variations detected by {\it Kepler} are dominated by a signal at a period of $938.507$~s, along with its first harmonic. In addition, a second modulation with a period of $953.262$~s is seen. The lightcurve, folded on the $938.507$~s period, shows a shape which is very similar to the superhump waveform found in AM\,CVn. In this picture the $953.262$~s modulation corresponds to the orbital period, whereas the $938.507$~s modulation is the superhump period. 

In AM\,CVn itself \citet{ski99} were able to explain all photometric periods in terms of only 3 basic periods that correspond to the orbital period and two additional periods, most likely due to disc precession. F11 detected 11 periods in the {\it Kepler} lightcurve of SDSS\,J1908, and could also explain all 11 periods in terms of only 3 basic periods. However, in that picture the $938.507$~s modulation corresponds to the orbital period, in contradiction to finding that the waveform at that period looks similar to the superhump waveform in AM\,CVn. Hence, the orbital period of SDSS\,J1908 remains ambiguous. 

A spectroscopic identification of coherent radial velocity changes is generally accepted as the most direct way of establishing the orbital period of the system. This formed the broad motivation for the present study. Additionally, the {\it Kepler} satellite kept SDSS\,J1908 as a short cadence target over the quarters Q6 to Q17, a total of 3 years: an unprecedented data set for any ultracompact binary.



\section{Observations and Data reduction}{\label{sec:observ}}
\subsection{Photometry}
We used the Q6 to Q17 short cadence data obtained by {\it Kepler} with a time resolution of 58.9\,sec. The original pixel data were downloaded from the Kepler Data Archive\footnote{http://archive.stsci.edu/kepler/}, resulting in 1.3\,million images in the {\it Kepler}-band which covers a total baseline of 1052\,days. The first observations were done on June 24th 2010 and the last were done on May 11th 2013.

There is a star only 5 arcsec away which contaminates the extracted flux of SDSS\,J1908 in the standard-pipeline. F11 identified this star as a G-star. We used point spread function (PSF) fitting as implemented in the PyKE tools provided by the NASA Kepler Guest Observer Office \citep{sti12} to separate the lightcurves of SDSS\,J1908 and the G-star. To correct for a linear trend caused by the instrument the lightcurve was normalised on a quarterly basis with a first order polynomial fit.  

\subsection{Spectroscopy}
We obtained phase-resolved spectroscopy of SDSS\,J1908 over five nights on 3 -- 7 July 2011 using the William Herschel telescope (WHT) and the ISIS spectrograph \citep{car93}. The full data set consists of 1875 spectra taken with the R1200B grating for the blue arm and the R1200R grating in the red arm covering a wavelength range of 4300 - 5070\,\AA\ and 5580 - 6244\,\AA\,respectively. All observations were done with a 1 arcsec slit and 2$\times$1 binning with the binning of 2 in the spatial direction. This resulted in a full-width half-maximum (FWHM) resolution of 0.92\,\AA\ for the R1200B grating and 1.04\,\AA\ for the R1200R grating. Each night an average bias frame out of 20 individual bias frames was made and a normalised flatfield frame was constructed out of 20 individual lamp flatfields. CuNeAr arc exposures were taken every hour to correct for instrumental flexure. Each exposure was wavelength calibrated by interpolating between the two closest calibration exposures. A total of 36 lines could be well fitted in each arc exposure using a Legendre function of order 4 resulting in 0.015\,\AA\ root-mean-square residuals. The best obtained signal-to-noise ratio (SNR) per spectrum was 4, whereas during July 5 we suffered from poor weather conditions and obtained only a SNR=1 per spectrum. This leads to a SNR of $\sim$70 per pixel in the grand average spectrum. BD+28\,4211 \citep{oke90} was used as a spectrophotometric standard to flux calibrate the spectra and correct them for the instrumental response. 

SDSS\,J1908 was observed over three nights on 15, 16 and 17 September 2011 using the GTC and the OSIRIS spectrograph \citep{cep98}. All observations were done with an 0.8 arcsec slit and the R2000B grating which covers a wavelength range of 3955 -- 5690\,\AA. Every exposure was binned 2$\times$2 on chip. This set-up results in a full-width half-maximum (FWHM) resolution of 2.65\,\AA. We obtained, each night, 20 bias frames to construct an average bias frame and 10 individual tungsten lamp flatfields to construct a normalised flatfield. A XeNe lamp spectrum was obtained at the beginning of the run as a master arc. About 15 lines were fitted using a Legendre function of order 5, resulting in an 0.07\,\AA\ root-mean-square residual. To save observing time, during the night about every 1.5 \,hr an Hg lamp spectrum was taken. This lamp has only three lines in the covered range but is good enough to account for small shifts of the spectra during the nights.  Additionally, the wavelength calibration for each individual spectrum was refined using sky lines. The average SNR of the individual spectra was around 30 (for a 60\,s exposure), resulting in a grand average spectrum with a SNR $>$200 per pixel. To correct for the instrumental response L1363--3 \citep{oke74} was used as a spectrophotometric standard. {\sc Molly}\footnote{{\sc Molly} was written by TRM and is available at http://www.warwick.ac.uk/go/trmarsh/software/} and {\sc Iraf}\footnote{{\sc Iraf} is distributed by the National Optical Astronomy Observatories, which are operated by the Association of Universities for Research in Astronomy, Inc., under cooperative agreement with the National Science Foundation} routines were used to reduce the data obtained with the WHT as well as the GTC. 

To obtain both high resolution and high SNR spectra, SDSS\,J1908 was also observed over one night on 12 July 2012 using Keck and the ESI spectrograph in Echellette mode ($R=\frac{\lambda}{\Delta\lambda}=8000$). The full dataset consists of 203 spectra. All observations were done with a 1.0 arcsec slit. An average flatfield frame was made out of 100 individual flatfield frames. CuAr arc exposures were taken every hour to correct for instrumental flexure. Each exposure was wavelength calibrated by interpolating between the two closest calibration exposures. {\sc Makee}\footnote{http://www.astro.caltech.edu/$\sim$tb/ipac\_staff/tab/makee/} was used to reduce the data. The SNR of the individual spectra was found to be around 6 (for a 60\,s exposure), resulting in a grand average spectrum with a SNR around 50 per pixel.

Additionally we took 4 spectra on 1 June 2014 of SDSS\,J1908 using Keck/ESI in Echellette mode with an exposure time of 15\,min each, for the spectroscopic analysis of the average spectrum. An average flatfield frame was made out of 10 individual flatfield frames. A HgNeXeCuAr lamp spectrum was obtained at the beginning of the night as a master arc. {\sc Makee} was used to reduce the data. The 4 spectra in combination with the data taken on 12 July 2012 result in a SNR of about 110 per pixel for the grand average spectrum.

Table\,\ref{observ} gives an overview of all observations and the instrumental set-ups. 

\section{Methods}
\subsection{Spectroscopic period determination}\label{sec:period_det}
To determine the orbital period the violet-over-red method (V/R) described in \citet{nat81} was used for the spectra, following \citet{roe05,roe06,roe07a} and \citet{kup13}. To maximise the SNR the ratios of the strongest helium absorption lines (4387\,\AA, 4471\,\AA, 4921\,\AA\ and 5015\,\AA) were summed. Lomb-Scargle (LS) periodograms of the measured violet-over-red-ratio as a function of the barycentric date were computed. We note that all times and phases in this analysis are in barycentric times.

The uncertainty on a derived period was estimated using the bootstrap method. In a simple Monte Carlo simulation 1000 periodograms were computed and in each the highest peak was taken as the orbital period (see \citealt{kup13}). A number of 369 spectra were randomly picked out of the full sample of 369 spectra, allowing for a spectrum to be picked more than once. The standard deviation on the distribution of the computed orbital period is taken as a measure of the accuracy in the derived period.

\begin{figure*}
\begin{center}
\includegraphics[width=0.99\textwidth]{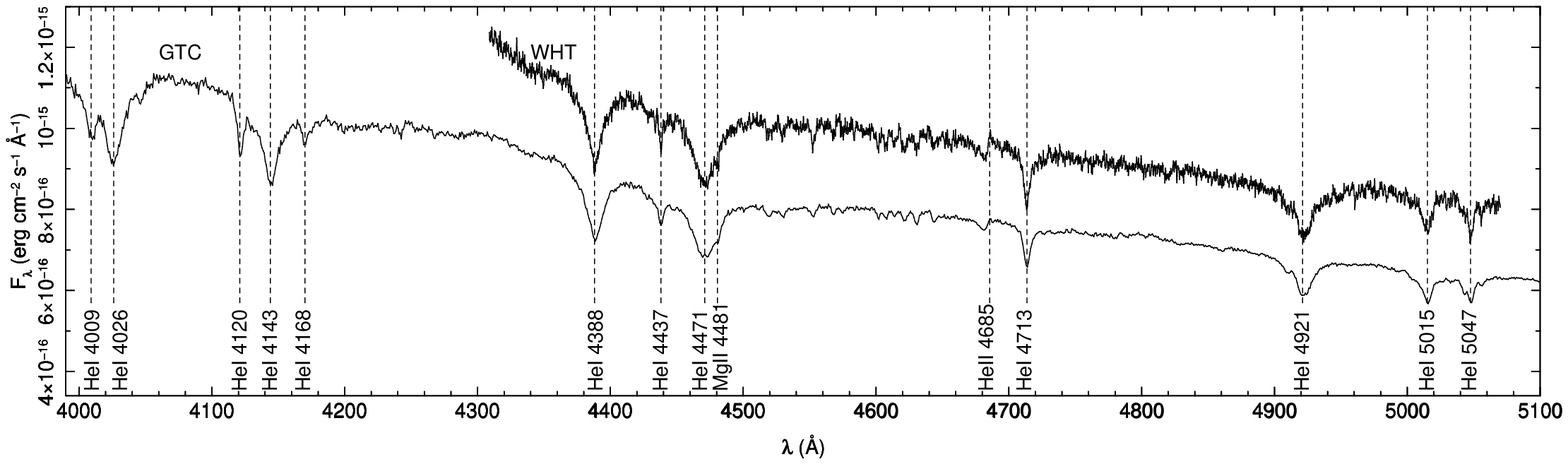}

\includegraphics[width=0.99\textwidth]{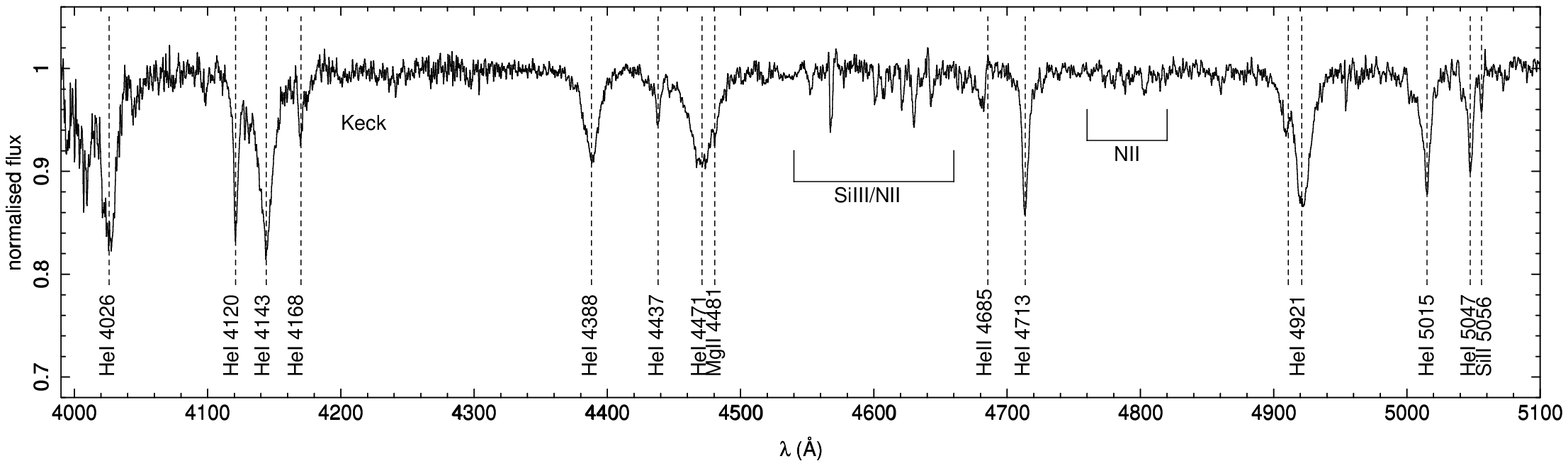}
\includegraphics[width=0.99\textwidth]{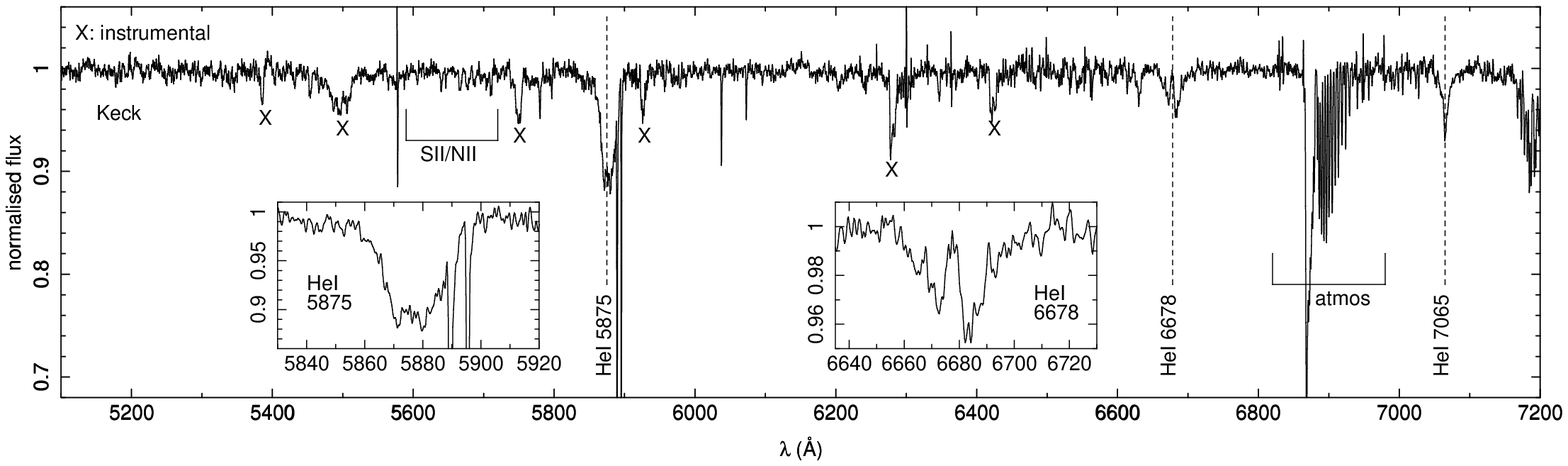}
\caption{Average spectrum of SDSS\,J1908 obtained with the WHT, Keck and GTC. Helium absorption lines of helium are indicated. {\bf Upper panel:} Average spectra obtained by the WHT and the GTC. {\bf Middle and lower panel:} Gaussian smoothed normalised average spectrum obtained by Keck.}
\label{whtaver}
\end{center}  
\end{figure*}

\subsection{Doppler tomography}\label{sec:doppler_tomo}
In Doppler tomography \citep{mar88} phase-resolved spectra are projected onto a two-dimensional map in velocity coordinates. We refer to \citet{ste03} and \citet{mar01} for reviews of Doppler tomography. Emission features that are stationary in the binary frame add up constructively in a Doppler tomogram while emission that is not stationary in the binary frame or moves on a period different from the orbital period will be spread out over the Doppler tomogram. Therefore, Doppler tomograms are useful to separate out features that move with a different velocity and/or different phase (e.g. bright spot and central spike). In this analysis Doppler tomograms  were computed using the software package {\sc Doppler}\footnote{{\sc Doppler} was written by TRM and is available at http://www.warwick.ac.uk/go/trmarsh/software/} and were used to measure the systematic velocity and the velocity amplitudes of the individual lines in SDSS\,J1908. Absorption features were inverted to appear as emission lines in the Doppler tomogram analysis. 

To measure the systematic velocity of the individual lines we followed the approach introduced by \citet{roe05}. For a given trial wavelength, a feature in the spectra will appear blurred in a Doppler tomogram if the "rest" wavelength does not coincide with the trial wavelength. We thus make Doppler tomograms for a range of trial wavelengths around the rest-frame wavelength of the spectral lines and fit a 2D Gaussian to the emission feature in every Doppler tomogram. For each line the height of the fitted spot peaks strongly around a certain wavelength. The maximum of a parabolic fit to the peak heights defines the "rest" wavelength and therefore the systemic velocity. In the next step Doppler tomograms accounting for the systemic velocity were computed. The center of a 2D Gaussian fit was calculated for each line which was taken as the position of the spot.   

\begin{figure*}
\begin{center}
\includegraphics[width=0.31\textwidth]{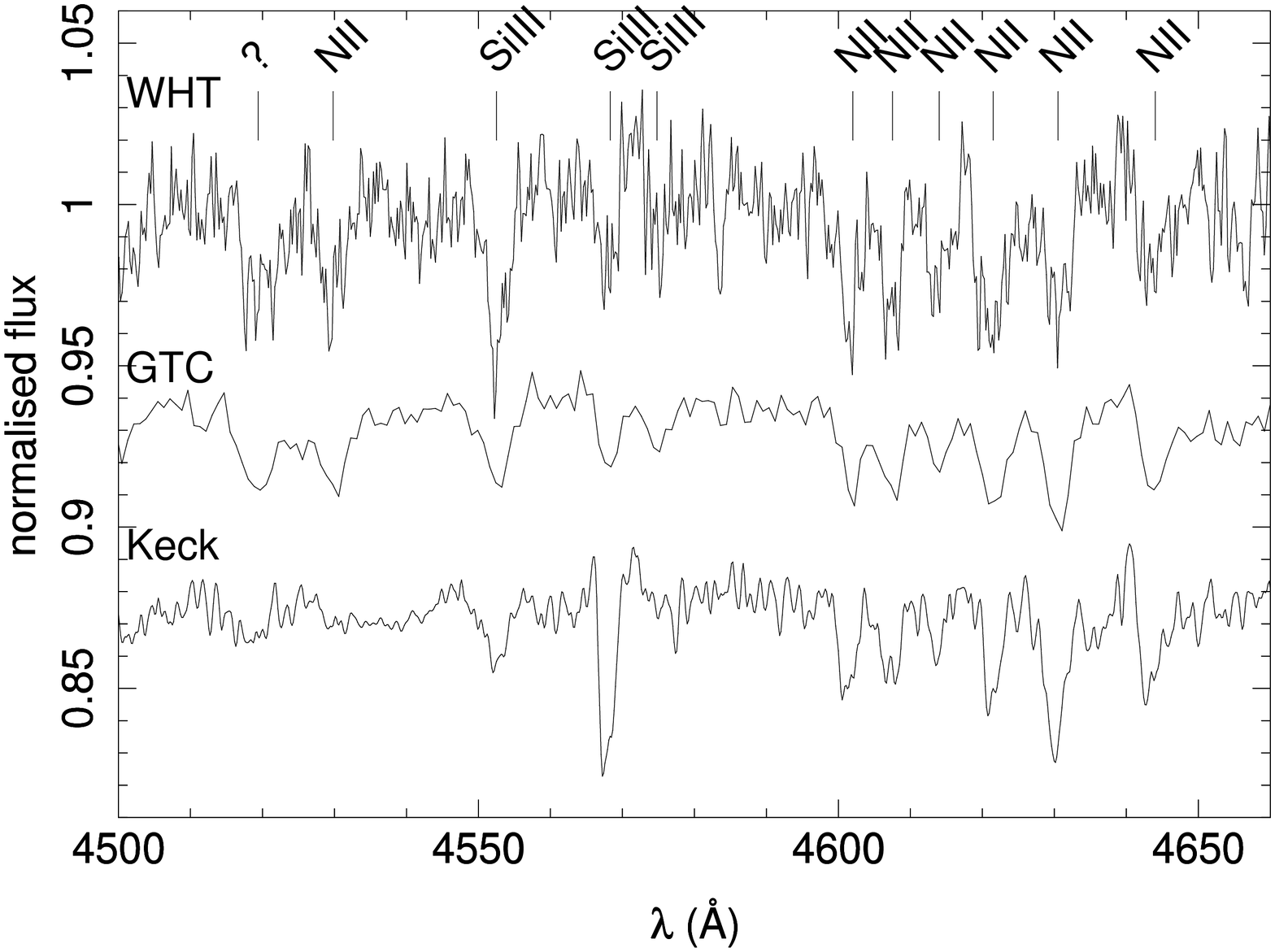}
\hspace*{0.01cm}
\includegraphics[width=0.31\textwidth]{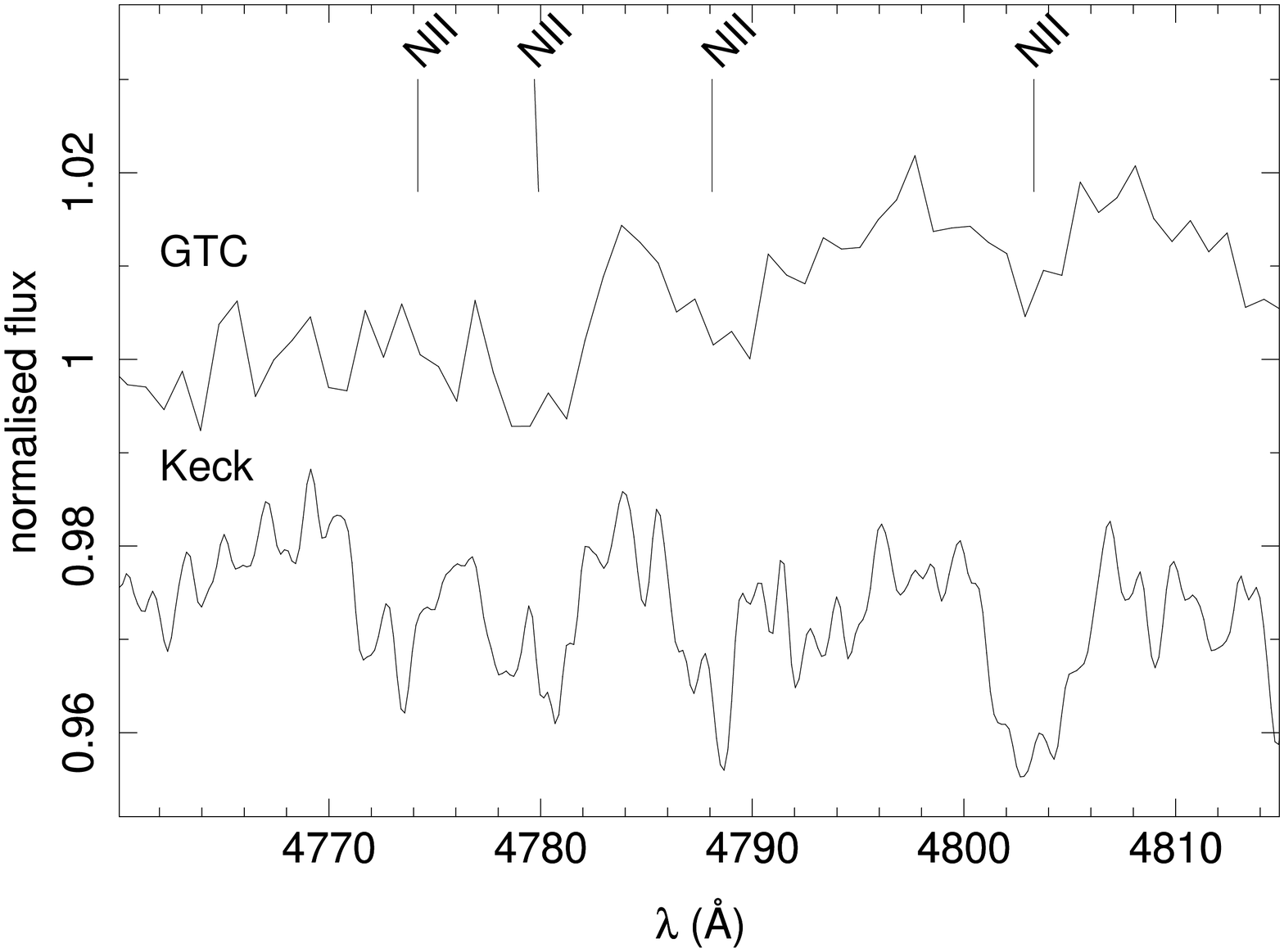}
\hspace*{0.01cm}
\includegraphics[width=0.31\textwidth]{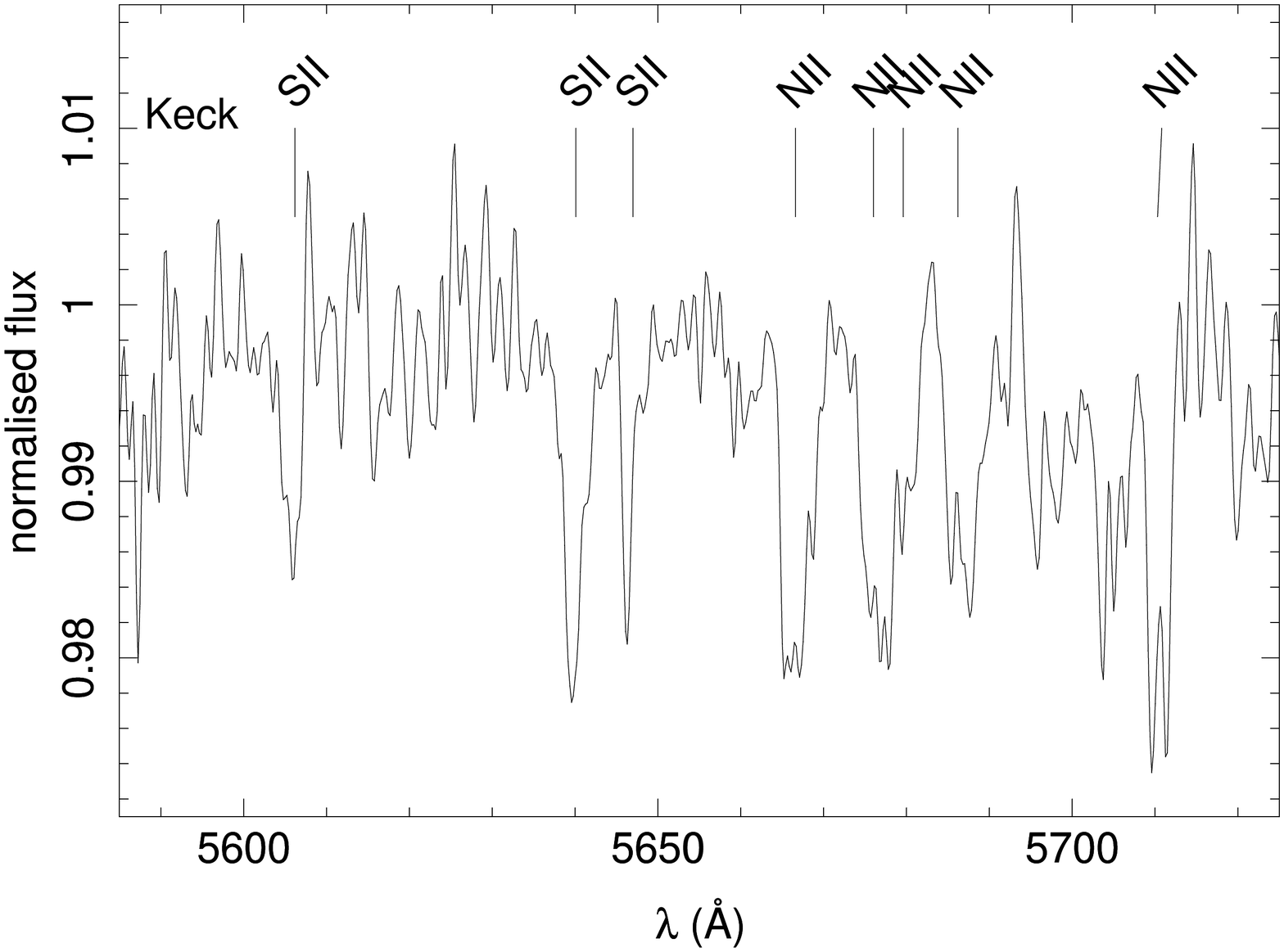}
\caption{Zoomed region of the average WHT, GTC  and Keck spectra around locations of various metal lines (N\,{\sc ii}, Si\,{\sc iii} and S\,{\sc ii}).}
\label{whtmetal}
\end{center}  
\end{figure*}

\subsection{Analysis of the {\it Kepler} lightcurve}\label{sec:ocanal}
To derive the periods seen in the {\it Kepler} lightcurve, a Fourier transform was computed using the tool {\sc Period04}\footnote{https://www.univie.ac.at/tops/Period04/}. For dynamical lightcurve analyses the discrete Fourier transform (FT) for the {\it Kepler} data set of our three lightcurve window were computed using a 200 days sliding window of the data. A block of 200 days of data has to be used to show a significant signal for some of the weaker periods in the FT.  

Observed-minus-computed (O-C) diagrams are a powerful tool that can be used to refine the periods and search for period variations. These diagrams compare the timing of an event, which in our case is the time of phase zero in the lightcurve based on an ephemeris (observed), to when we expect such an event if it occurred at an exactly constant periodicity (computed). A linear trend in an O-C diagram corresponds to an incorrect period, whereas a parabolic trend is caused by a period derivative. 

O-C diagrams were computed for the 5 strongest frequencies detected in SDSS\,J1908 (including the frequencies identified as possible orbital periods by F11), as well as the orbital period to search for period variations. For the three stronger periods ($90, 92$ and $184$\,c/d) two-week blocks of data were folded on a fixed period with a fixed zero-point. For the less strong periods ($74, 79$ and $111$\,c/d) two month blocks of data were folded on a fixed period with a fixed zero-point. 

\begin{figure}
\begin{center}
\includegraphics[width=0.35\textwidth, angle=270]{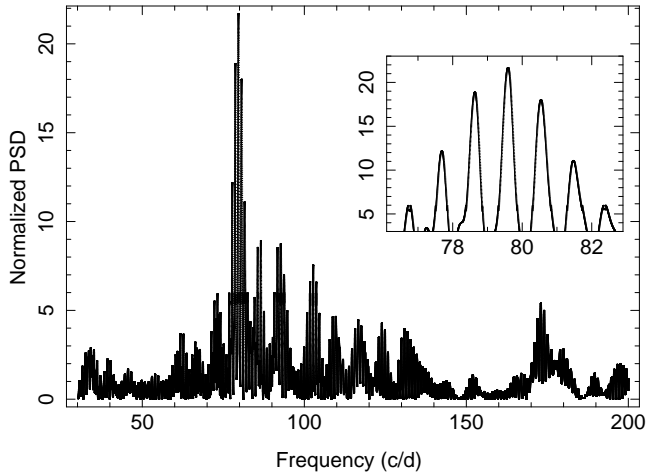}
\caption{LS periodograms of the GTC data using the red over violet wing flux ratio. The strongest peak was chosen to be the correct period.} 
\label{fig:ls_spec}
\end{center}
\end{figure} 

For the frequencies which show a sinusoidal variation in the phase folded lightcurve ($90, 92, 111, 184$\,c/d), a sine curve was fitted to the folded lightcurve to obtain the phasing. For the other two frequencies ($74, 79$\,c/d) a Gaussian was fitted to the phase folded lightcurve and phase of the center of the maximum was measured. In the next step, we moved one week forward in time and repeated the procedure to measure the phase. This was done for the full $1052$\,days of data. If the correct period is used no linear trend of the measured phase is expected. Therefore, a constant was fitted to each O-C diagram and the $\chi^2$ value of the fit was computed. To find the best period a parabola was fitted to the computed $\chi^2$ value and the minimum was assumed to be the best period. See Sec.\,\ref{sec:ocdiag} for the full discussion on the O-C diagrams of the 5 strongest frequencies.


Note that the Kepler lightcurve is given in barycentric Julian date (BJD). Therefore we computed BJDs for the spectroscopic data which allows us to compare the spectroscopic and the photometric datasets.

\begin{table}
 \centering
 \caption{Velocities of the emission and absorption lines seen in SDSS\,J1908}
  \begin{tabular}{cccc}
  \hline
  Feature         &  $\gamma$ & $K_x$  & $K_y$ \\
                  &  (km\,s$^{-1}$)  &  (km\,s$^{-1}$)  &  (km\,s$^{-1}$) \\
                     \hline\hline
   Helium absorption lines  &                  &                   &                 \\     
   He\,{\sc i} 4387/4921    &  23.7$\pm$4.4    &       -           &       -         \\
   He\,{\sc i} 4713         &  25.0$\pm$3.8    &    116.8$\pm$6.5  &    --23.3$\pm$4.5 \\ 
    \smallskip
   He\,{\sc i} 5047         &  10.1$\pm$5.0    &    84.8$\pm$9.5  &    --35.3$\pm$7.3 \\ 
   Helium emission lines    &                  &                   &                 \\
  \smallskip
   He\,{\sc i} 4471         &  15.6$\pm$5.3    &    --111.2$\pm$6.8  &   36.8$\pm$6.6 \\
   Metal absorption lines   &                  &                   &                 \\
   N\,{\sc ii} 4601 - 4643  &  2.3$\pm$5.0     &       -           &       -         \\
   Si\,{\sc iii} 4552/4567/4574 & 2.7$\pm$4.6  &       -           &       -         \\
   Mg\,{\sc ii} 4481        &  1.0$\pm$5.5     &     109.6$\pm$9.3 &    5.6$\pm$7.1 \\ 
   Si\,{\sc ii} 5041/5055   &  6.2$\pm$5.8     &     100.0$\pm$7.2 &    --10.4$\pm$7.6 \\  
       \hline
\end{tabular}
\label{tab:vel}
\end{table}

\section{Results}
\subsection{Average spectra}
\subsubsection{Helium absorption lines}
The average spectra of SDSS\,J1908 are shown in Fig.~\ref{whtaver}. The strong absorption lines of neutral helium are clearly visible. Helium absorption lines are seen in AM\,CVn systems during dwarf-novae type outbursts, as well as in high state systems with orbital periods $<$\,$20$\,min (e.g. \citealt{roe07, lev13}). SDSS\,1908 appears very similar to the short period systems AM\,CVn and HP\,Lib \citep{roe06a, roe07}. The He\,{\sc i} 6678 and 5875\,\AA\ line shows an emission core which is also observed in AM\,CVn itself \citep{pat93}. The helium emission lines in AM\,CVn and HP\,Lib are much broader than in SDSS\,J1908, indicating that the lines of SDSS\,J1908 show less rotational broadening and the system is seen under lower inclination, as also concluded by F11.  

\begin{figure*}
\begin{center}
\includegraphics[width=0.99\textwidth]{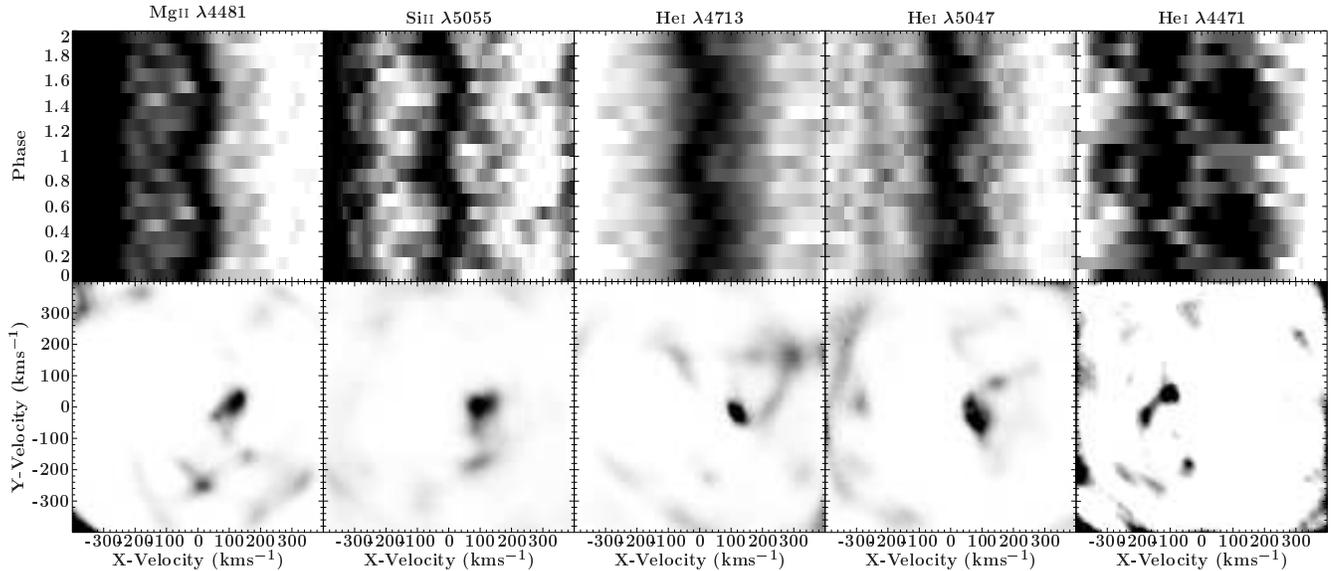}
\caption{Trailed spectra (top row) and maximum-entropy Doppler tomograms (bottom row) of selected metal and He\,{\sc i} lines of SDSS\,J1908. The Si\,{\sc ii}, Mg\,{\sc ii} and He\,{\sc i} 4713/5047\,\AA\,absorption lines and the He\,{\sc i} 4471\,\AA\, emission core is shown. Visible is a phase offset between the emission and absorption features. The average line profile has been divided out to enhance the visibility of the spot on the Doppler tomograms.} 
\label{fig:1908doppler}
\end{center}
\end{figure*} 

\begin{figure*}
\begin{center}
\includegraphics[width=0.99\textwidth]{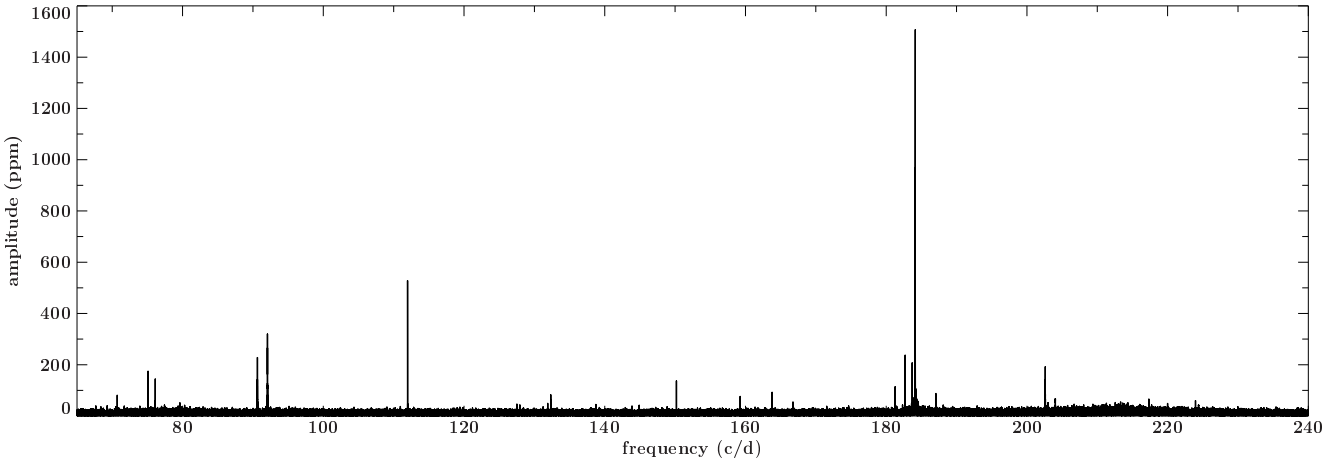}
\caption{Fourier transform of the lightcurve obtained with {\it Kepler} of SDSS\,J1908.} 
\label{fig:lomb_star}
\end{center}
\end{figure*}  


\subsubsection{Metal absorption lines}
SDSS\,J1908 shows a variety of weak metal lines including a number of nitrogen lines as well as silicon and sulphur lines (see Fig.\,\ref{whtmetal} for a selection of metal lines). In contrast to the large number of detected nitrogen lines, there is no evidence for oxygen. The strongest O\,{\sc i} line in the optical is the triplet at 7771-7775\,\AA\,and the strongest O\,{\sc ii} lines in the optical are 4649 or 4414\,\AA. The strongest C\,{\sc ii} line in the optical is C\,{\sc ii} 4267\,\AA. This line might be visible just above the noise level in the GTC data at an equivalent width of $<0.1$\,\AA. Tab.\,\ref{tab:equi} gives an overview of the detected lines with measured equivalent widths. 

\subsection{Kinematic analysis}
\subsubsection{Spectroscopic orbital period}
Using the method described in Sec.~\ref{sec:period_det} the LS periodogram shows a clear peak at $1085.7\pm2.8$\,sec ($79.58\pm0.20$c/d, $18.095\pm0.046$\,min; Fig.~\ref{fig:ls_spec}) in the spectroscopic data obtained with the GTC. This period is in excellent agreement with a period at $1085.10(9)$\,sec detected in the {\it Kepler} lightcurve (see Sec.~\,\ref{sec:ocdiag}).

To test whether this period is indeed the orbital period, the GTC and WHT spectra were phase folded on the {\it Kepler} period. Phase folded spectra and the Doppler tomograms were computed and analysed following Sec.~\ref{sec:doppler_tomo}. Phase folded spectra and Doppler tomograms are shown in Fig.\,\ref{fig:1908doppler}. We find a weak S-wave in emission in the core of He\,{\sc i} 4471\,\AA. The helium lines He\,{\sc i} 4713/5015/5047\,\AA, the Si\,{\sc ii} 5041/5055\,\AA\,and the Mg\,{\sc ii} 4481\,\AA\,lines show an S-wave in absorption. The phase offset between the emission lines and the absorption lines is found to be of $170\pm15^\circ$.

No line variation was detected for He\,{\sc i} 4387/4921\,\AA, Si\,{\sc iii} 4552/4567/4574\,\AA\,and N\,{\sc ii} 4601 - 4643\,\AA. However, the helium lines show a  different systemic velocity than the silicon and nitrogen lines. See Tab.\,\ref{tab:vel} for an overview of the measured velocities.  

Because the period at $1085.10(9)$\,sec detected in the {\it Kepler} lightcurve is not the most prominent period in the {\it Kepler} lightcurve, the GTC and WHT data were also folded on different periods seen in the Fourier transform of the {\it Kepler} data. However, no variations can be seen in any other period. We therefore conclude that $1085.10(9)$\,sec is the orbital period of SDSS\,J1908.



\subsection{Lightcurve variations}
We find no strong variation, such as dwarf novae type outbursts or superoutbursts, in the normalised lightcurve. The absence of strong variations is well in agreement with an AM\,CVn type binary with an orbital period below $20$\,min \citep{pat93,pat02}.
 

\begin{figure}
\begin{center}
\includegraphics[width=0.48\textwidth]{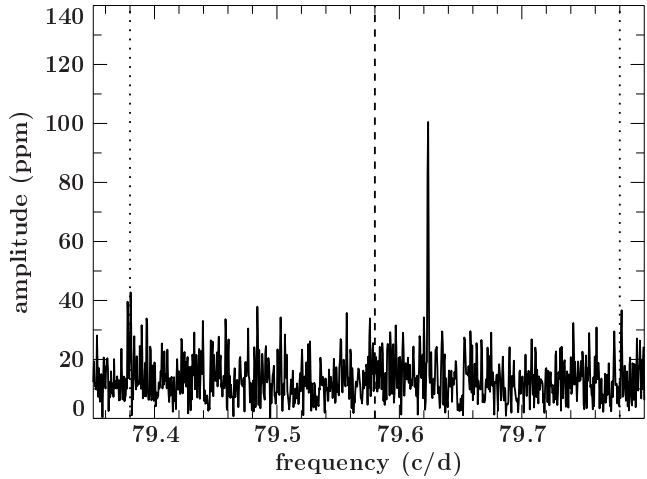}
\caption{View around the peak at $79.6233$\,c/d in the Fourier transform of the {\it Kepler} lightcurve. Shown is the position of the spectroscopically determined orbital period (dashed line) with its error (dotted lines).} 
\label{fig:strong_peak_79}
\end{center}
\end{figure}  

\begin{figure}
\begin{center}
\includegraphics[width=0.48\textwidth]{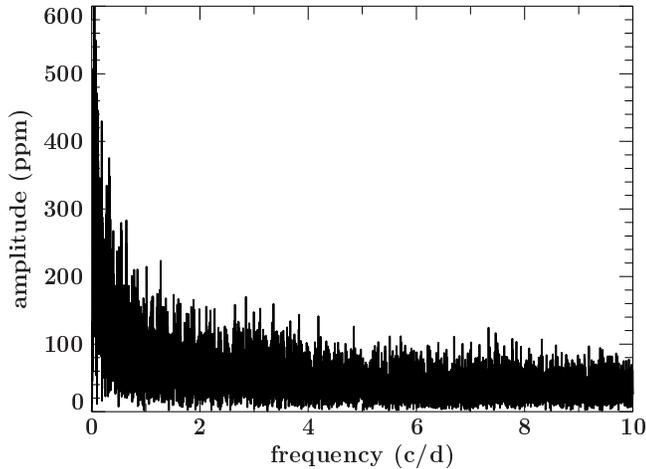}
\caption{Zoomed view on the low frequency region in the Fourier transform.} 
\label{fig:noise} 
\end{center}
\end{figure}  

To determine and refine the periods in SDSS\,J1908, and to search for possible periodic variations in the close-by G-star, we used the 1052\,days lightcurves obtained by the {\it Kepler} satellite {\rm for both objects}. Fourier transforms (FT) of the lightcurves for both objects were computed. No periodic variations were detected in the G-star over the frequency range $0$-$250$\,c/d. This shows that even if the photometric disentangling is not perfect no influence of the position or occurrence of periods detected in SDSS\,J1908 is expected to come from the G-star. 

Fig.\,\ref{fig:lomb_star} shows the FT of the full lightcurve for SDSS\,J1908. We use the mean value of the FT amplitude spectrum of the full data set to approximate the significance on the amplitudes in Fourier space, $\sigma_{\rm FT mean}=14$\,ppm. We adopt $4\sigma_{\rm FT mean}=52$\,ppm as the threshold of significance of the peaks in the FT amplitude spectrum. Several peaks ranging from periods of $196$\,sec up to $1221$\,sec can be identified in the FT of the full lightcurve. We find a total of $42$ signals well above the noise level in the FT amplitude spectrum (Tab. \ref{tab:periods}). The detected frequencies given in Tab. \ref{tab:periods} can only be seen as average frequencies over the observed period because the peaks show a frequency variability.  

\begin{figure}
\begin{center}
\includegraphics[width=0.48\textwidth]{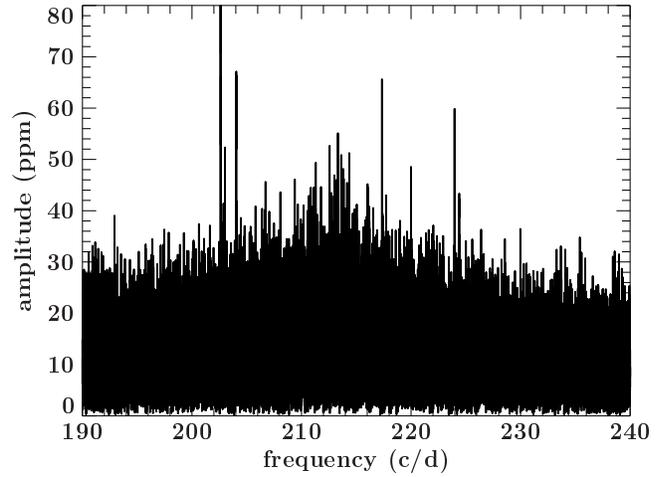}
\caption{Zoomed view around the QPO in the Fourier transform.} 
\label{fig:qpo}
\end{center}
\end{figure}  

\begin{figure}
\begin{center}
\includegraphics[width=0.48\textwidth]{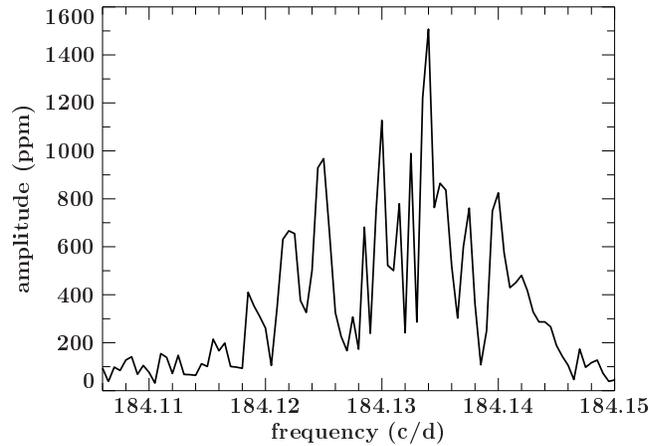}
\caption{Zoomed view around the strongest peak in the Fourier transform of the {\it Kepler} lightcurve.} 
\label{fig:peak}
 \end{center}
\end{figure} 

\begin{figure}
\begin{center}
\includegraphics[width=0.48\textwidth]{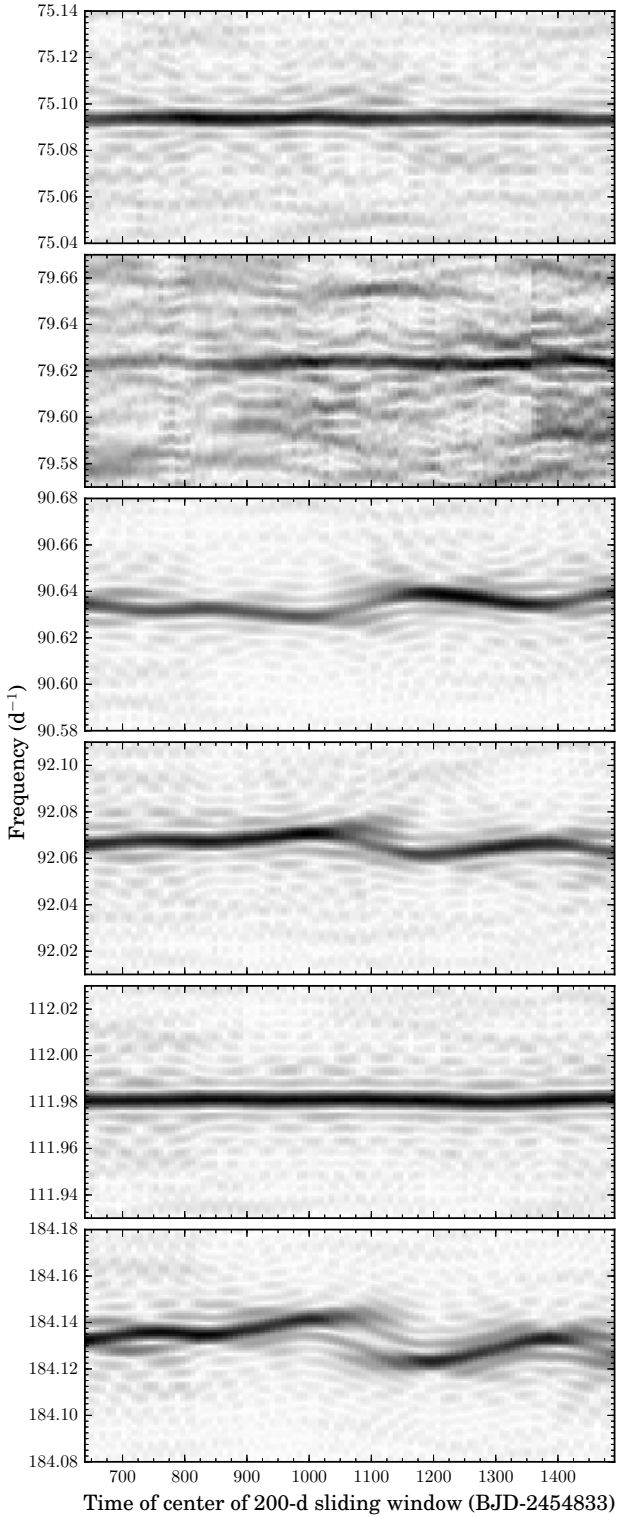}
\caption{Zoomed view of the two-dimensional discrete Fourier transform for the full data set derived from the {\it Kepler} lightcurve. Shown here is the amplitude vs. time and frequency for the orbital frequency and the five strongest frequencies detected in the lightcurve. We use a 200-day sliding window, and darker greyscale corresponds to higher amplitudes. } 
\label{fig:fourier_maps}
\end{center}
\end{figure}

For stable frequencies, pre-whitening the light curve by the highest peaks in a FT removes most power in that region and allows for a relatively simple extraction of the periods present in SDSS\,J1908. However, none of the peaks in the FT of SDSS\,J1908 can be pre-whitened in the standard way, because most frequencies show a variability in the frequency and therefore a pre-whitening of the strongest does not completely remove the pulse shape but leaves a new peak at slightly different frequency as the pre-whitened peak. See Sec.\,\ref{sec:ocdiag} for further discussion on the stability.


Fig.\,\ref{fig:strong_peak_79} shows the periodogram of the {\it Kepler} lightcurve at the position where the spectroscopic orbital period was detected. We find a single peak at $79.6234$\,c/d. The position of this peak is consistent with the orbital period derived from the spectroscopic data within one sigma. 

Besides the strong periodic variations some intriguing features can be seen in the Fourier transform which were also notified by F11. Below a frequency of about $20$\,c/d the noise is strongly increasing as shown in Fig.\,\ref{fig:noise}. This feature is not seen in the G-star and therefore expected to be a real feature in SDSS\,J1908 and not caused by the satellite. Those variations are due to low-frequency variations on timescales of hours to days and are well known from cataclysmic variables and from accreting low mass X-ray binaries as a sign of accretion \citep{kli05,sca14}. 

\citet{pat02} discovered a quasi-periodic oscillation (QPO) in HP\,Lib at a frequency of $280 - 320$\,c/d. We discover a similar QPO in SDSS\,J1908 at a frequency of about $200 - 230$\,c/d corresponding to a period of $6 - 7$\,min (see Fig.\,\ref{fig:qpo}).

\section{Periodic stability in the {\it Kepler} data} \label{sec:ocdiag}
We found a large number of frequencies which show several close-by peaks in the FT. This means that the system shows either several periods with similar stable frequencies or one single frequency shows frequency variability in time (Fig.\,\ref{fig:peak}). Figure\,\ref{fig:peak} shows the strongest variation detected in the lightcurve of SDSS\,J1908 at a frequency of $P=184.1326$\,c/d which shows a strong frequency variability.

Fig.\,\ref{fig:fourier_maps} shows the discrete Fourier transform amplitude spectra for the Kepler lightcurve data set, created following Sec.\,\ref{sec:ocanal}. We show the amplitude spectra for the spectroscopic orbital period as well as the five strongest peaks in the FT of the full lightcurve. 



For the first half year of observations the spectroscopic period is only marginally above the noise level with an amplitude of $59$\,ppm in the FT. After half a year the amplitude of the orbital period increases to $109$\,ppm and stays well above the noise level over the full observing period. During the last year of observations the signal strength reaches its maximum with an amplitude of $\sim150$\,ppm ($2$nd panel in Fig.\,\ref{fig:fourier_maps}). A variable strength in the signal of the orbital period in the FT is also observed in AM\,CVn itself \citep{ski99}. 

The frequencies around $75.09$\,c/d and $111.98$\,c/d ($1$st and $4$th panel in Fig.\,\ref{fig:fourier_maps}) are visible over the full observed period with a constant strength at $\sim330$\,ppm for $75.09$\,c/d and $\sim560$\,ppm for $111.98$\,c/d. Both periods show no significant frequency variations.

\begin{table}
 \centering
 \caption{Overview of the periods measured from O-C diagrams}
  \begin{tabular}{ccc}
  \hline
  Feature         &  Frequency (1/d) &  Period (sec)  \\
                     \hline\hline
 $P_{\rm 75}$      &  75.093(5)   &   1150.56(5)    \\
 $P_{\rm orb}$     &  79.623(3)   &   1085.10(9)   \\
 $P_{\rm 90}$      &  90.633(5)   &   953.29(0)    \\
 $P_{\rm 92}$      &  92.066(1)   &   938.45(6)   \\  
 $P_{\rm 111}$     & 111.980(5)   &   771.56(3)   \\
 $P_{\rm 184}$     & 184.132(6)   &   469.22(7)   \\  
       \hline
\end{tabular}
\label{tab:refine}
\end{table}

\begin{figure*}
\begin{center}
\includegraphics[width=0.495\textwidth]{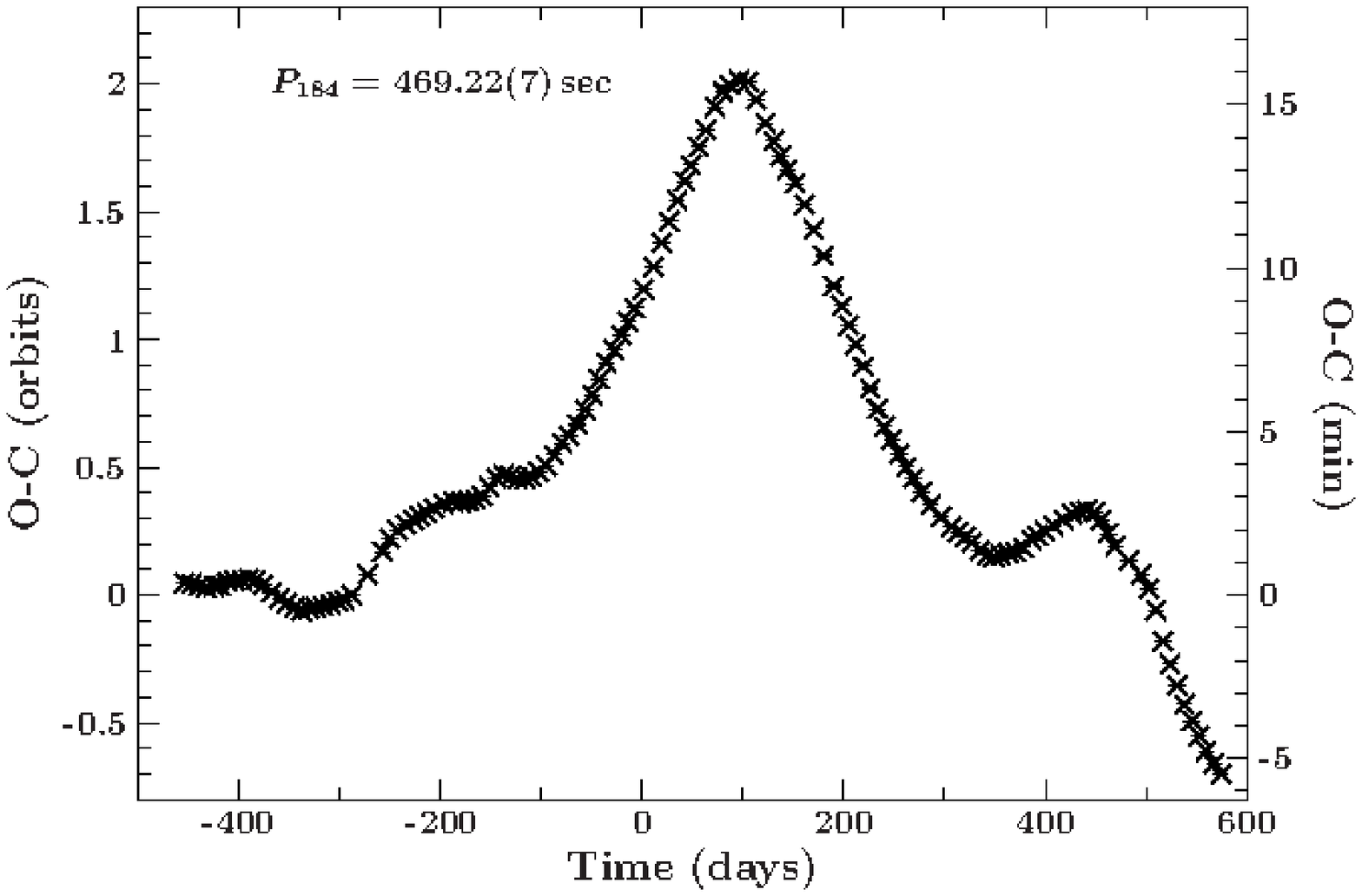}
\includegraphics[width=0.485\textwidth]{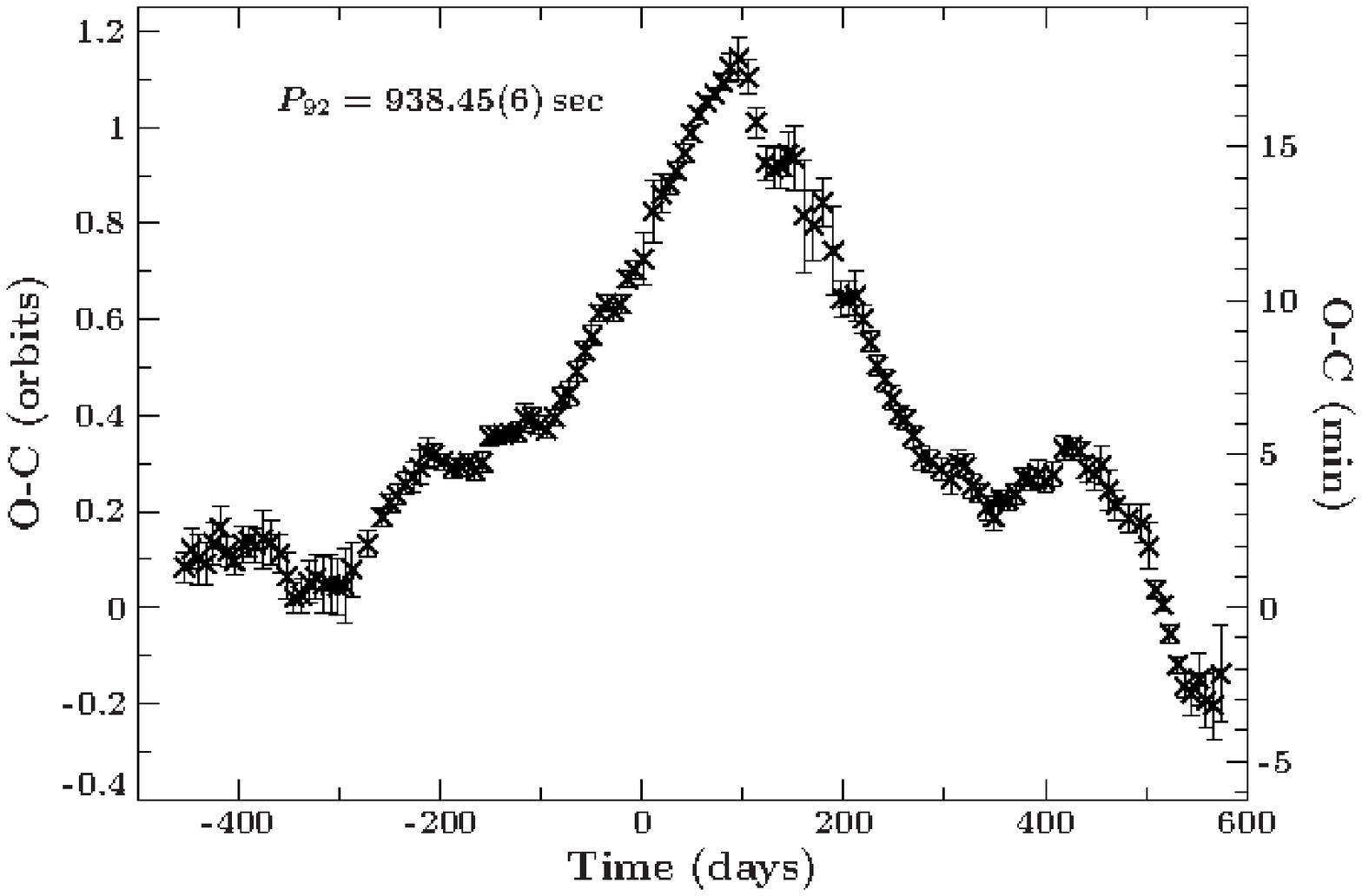}
\hspace*{0.05cm}
\includegraphics[width=0.485\textwidth]{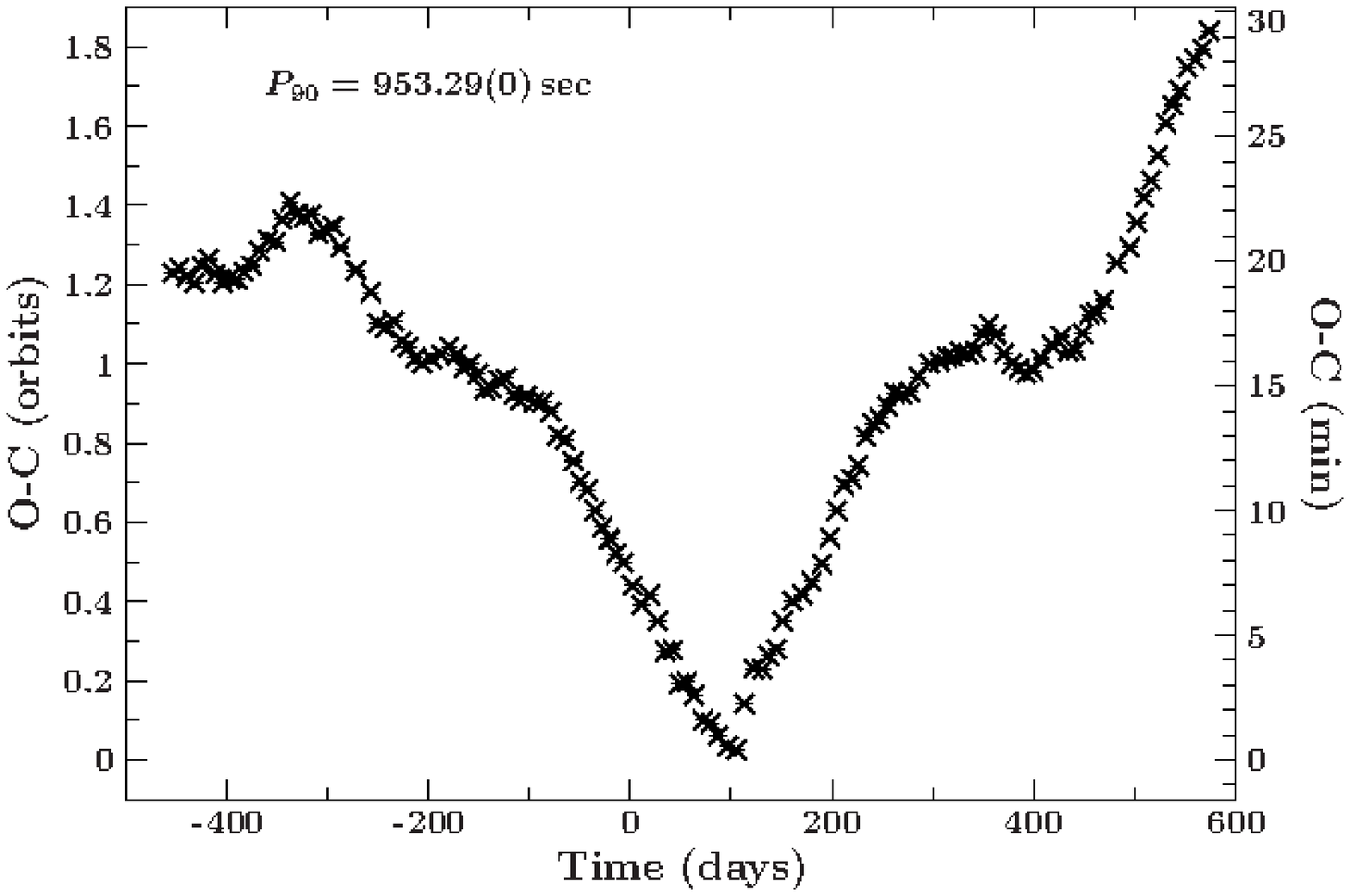}
\includegraphics[width=0.485\textwidth]{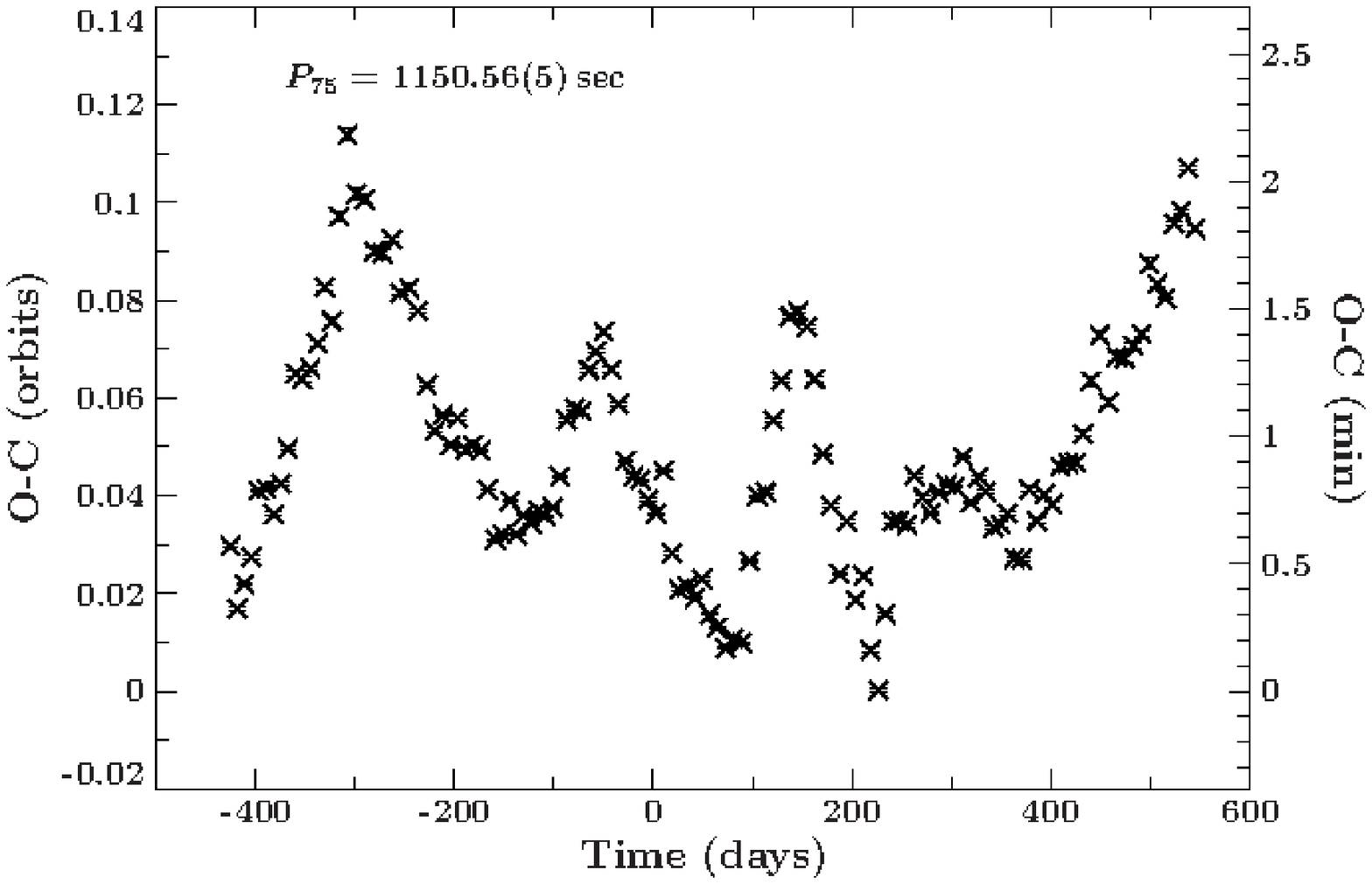}
\hspace*{0.05cm}
\includegraphics[width=0.485\textwidth]{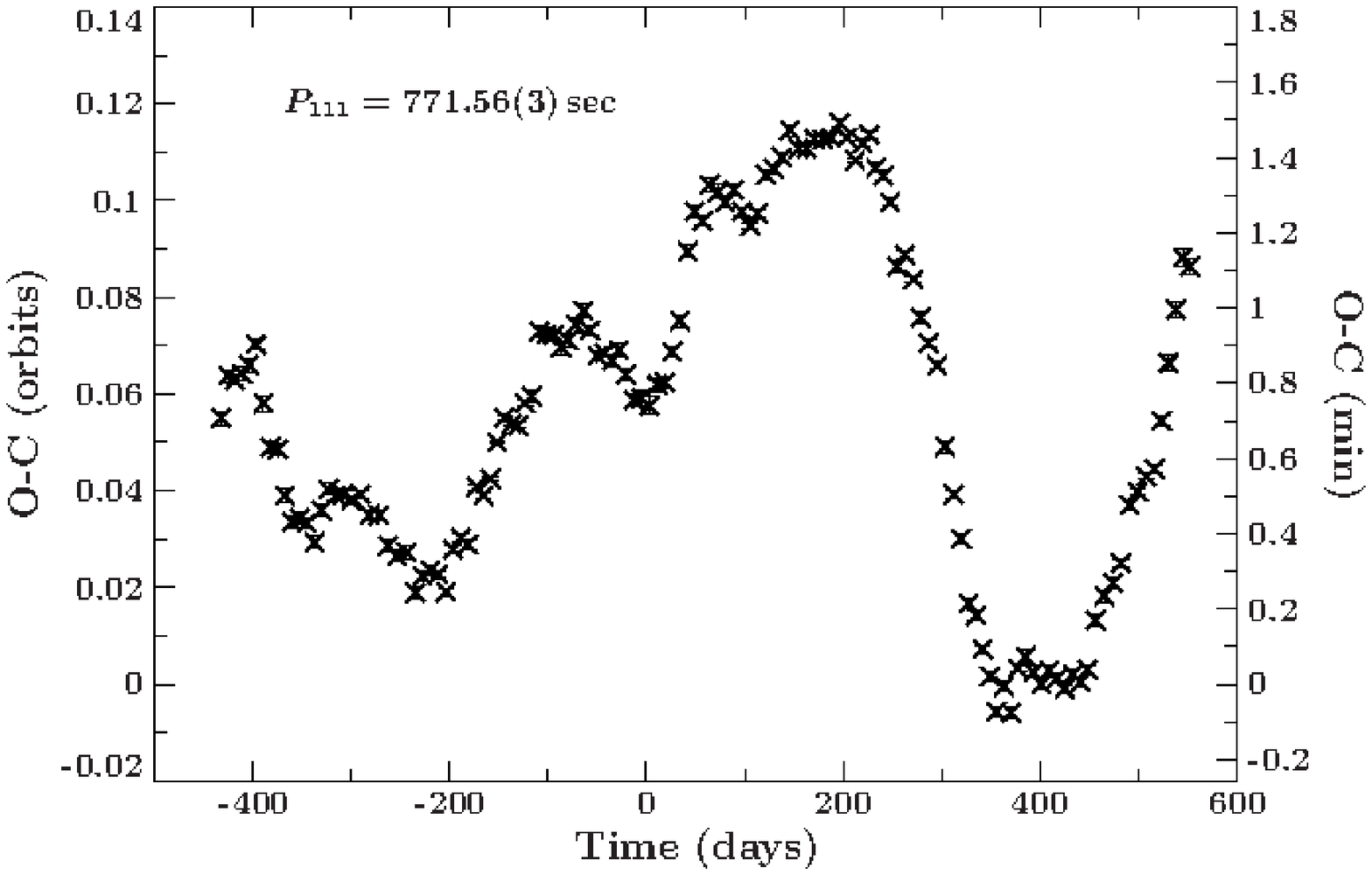}
\includegraphics[width=0.485\textwidth]{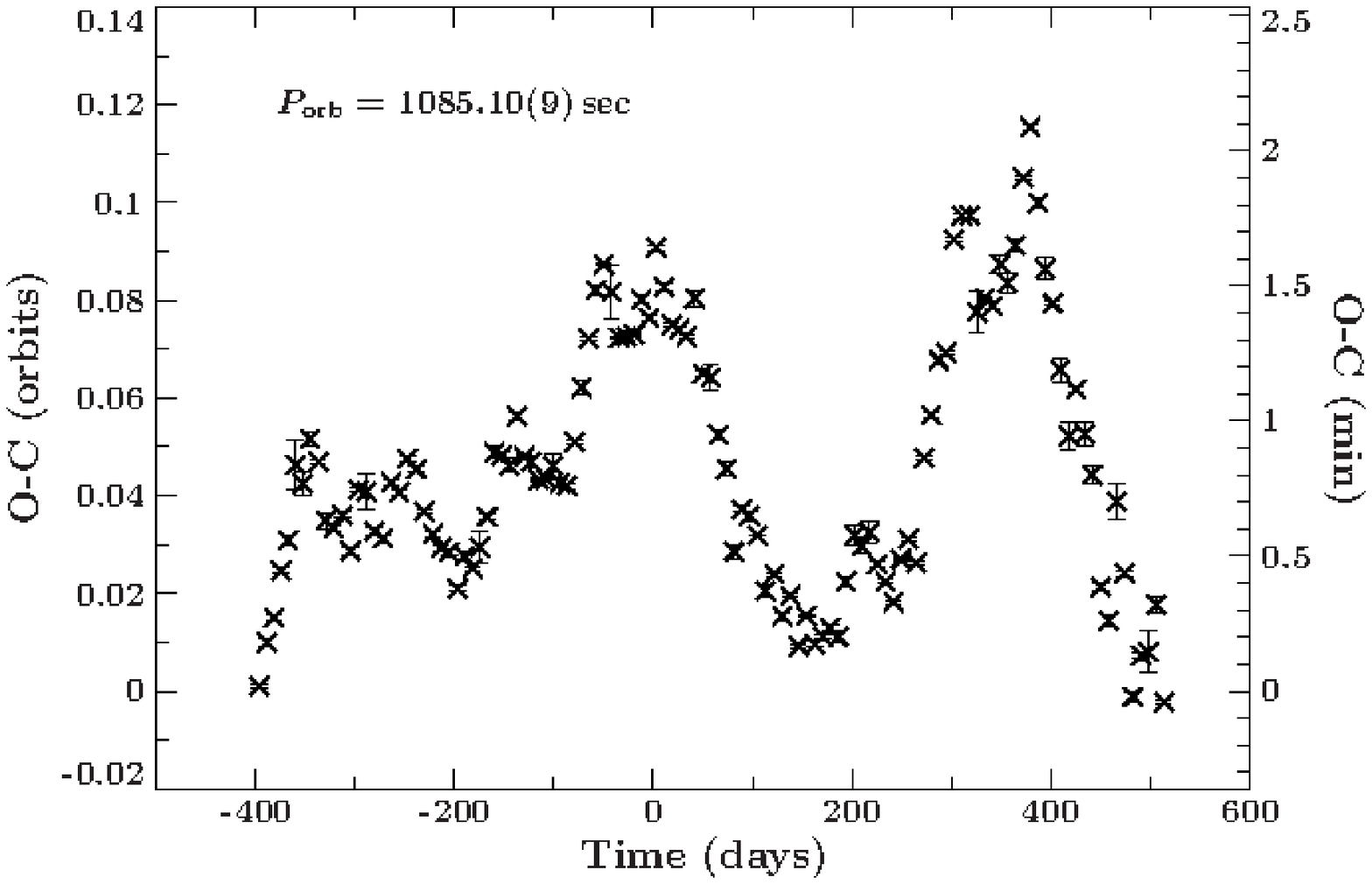}
\caption{O-C diagrams of the five strongest variations detected in the lightcurve as well as the spectroscopic orbital period.}
\label{fig:o-c}
\end{center}
\end{figure*}  

\begin{figure*}
\begin{center}
\includegraphics[width=0.485\textwidth]{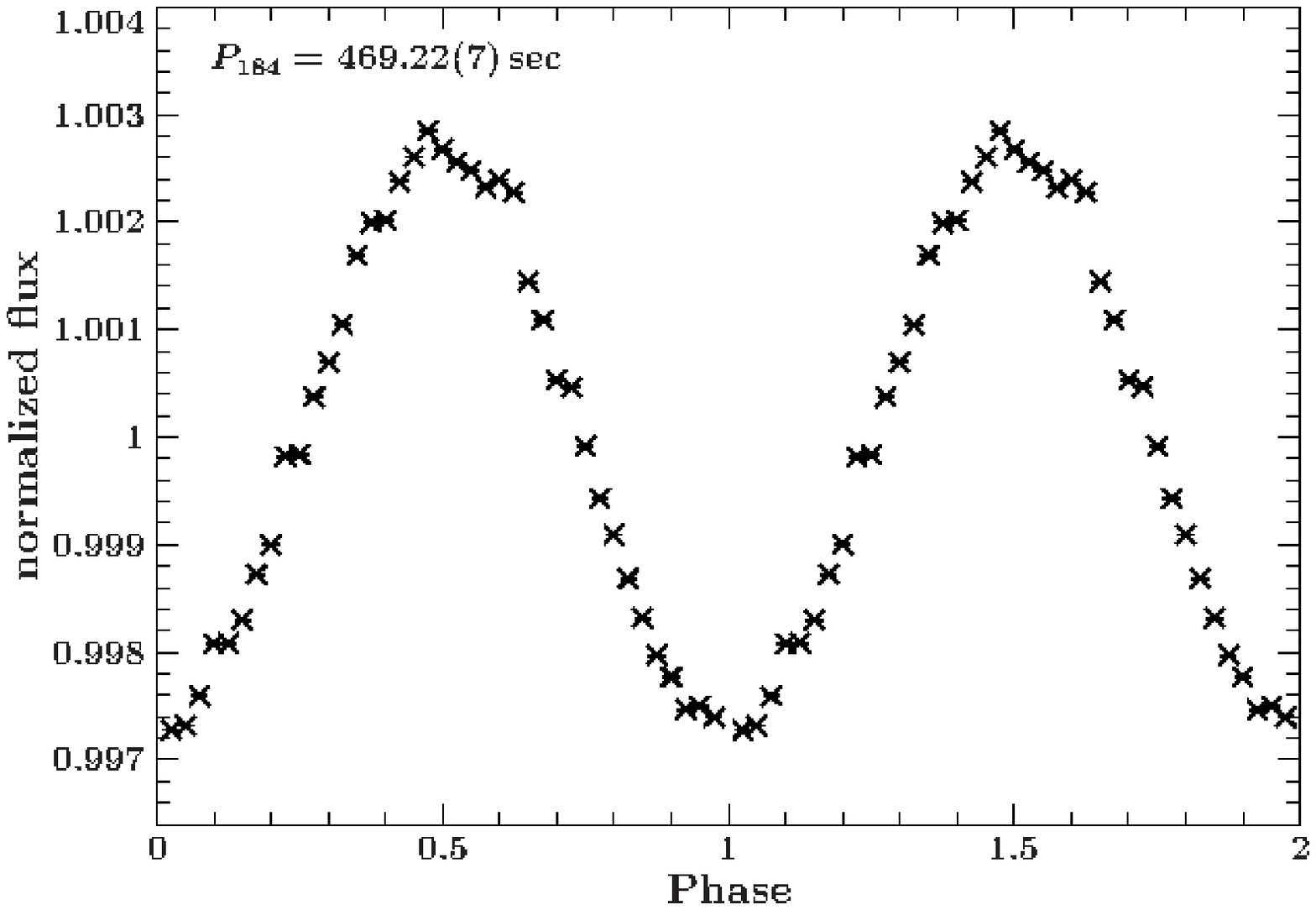}
\includegraphics[width=0.485\textwidth]{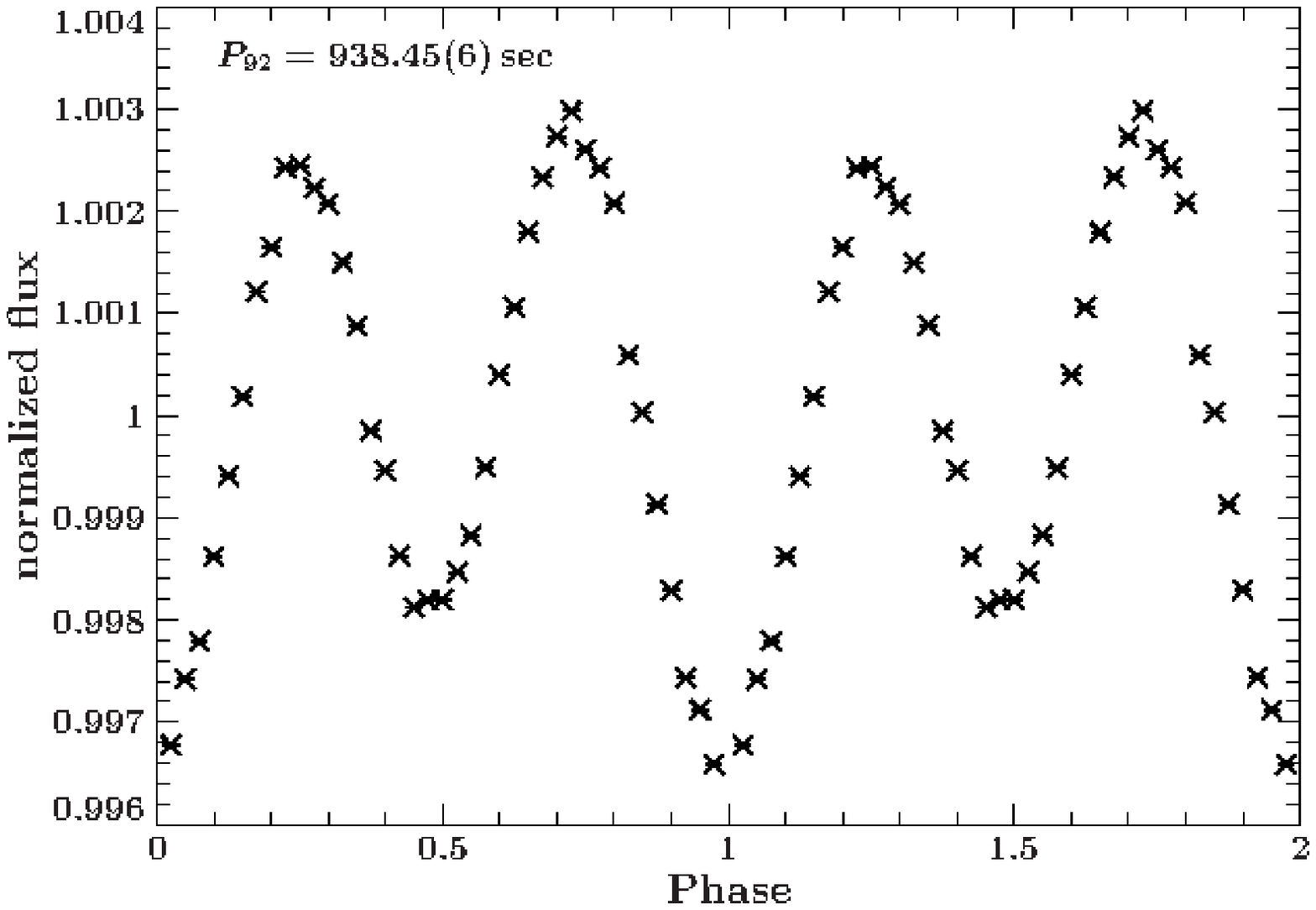}
\hspace*{0.05cm}
\includegraphics[width=0.485\textwidth]{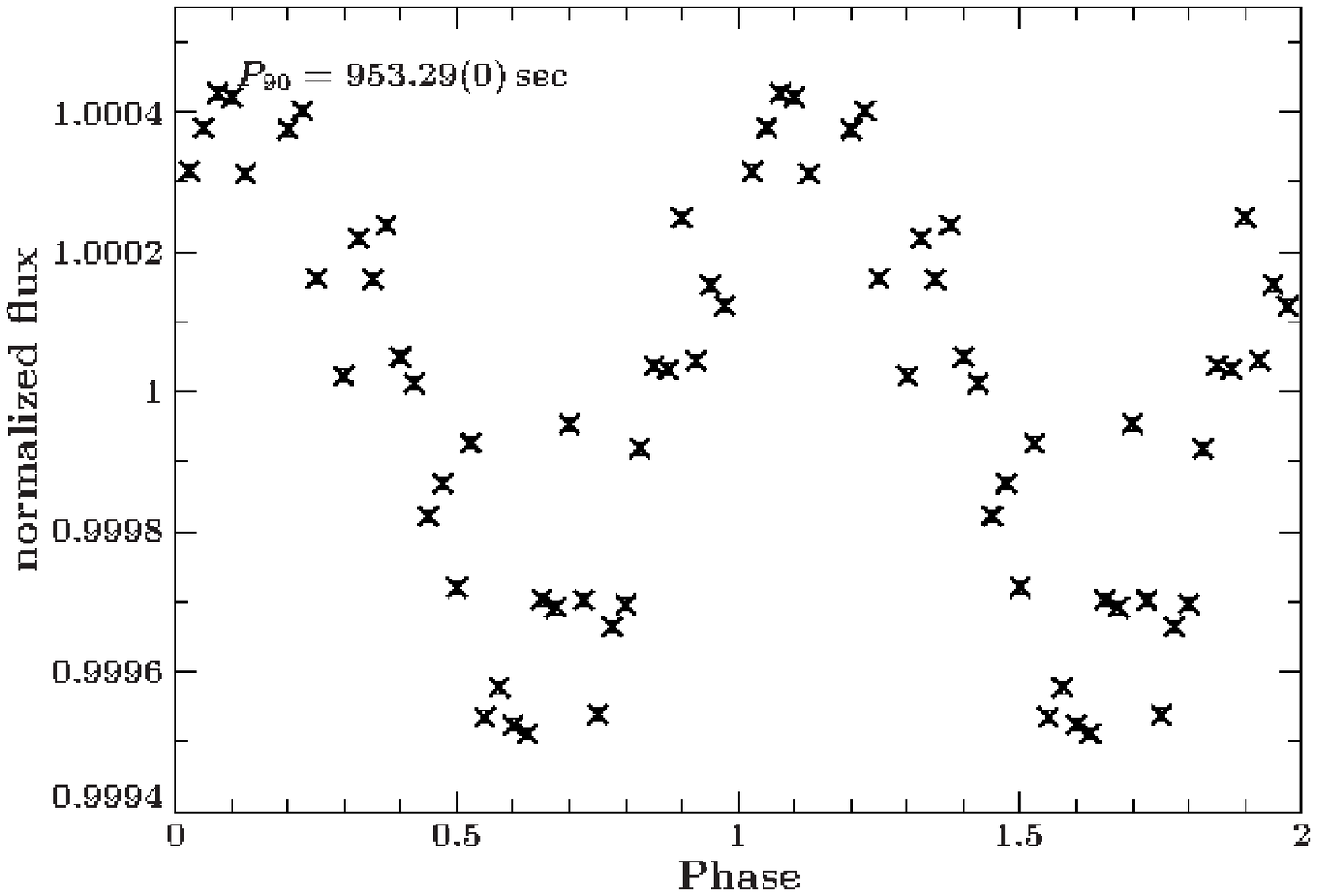}
\includegraphics[width=0.485\textwidth]{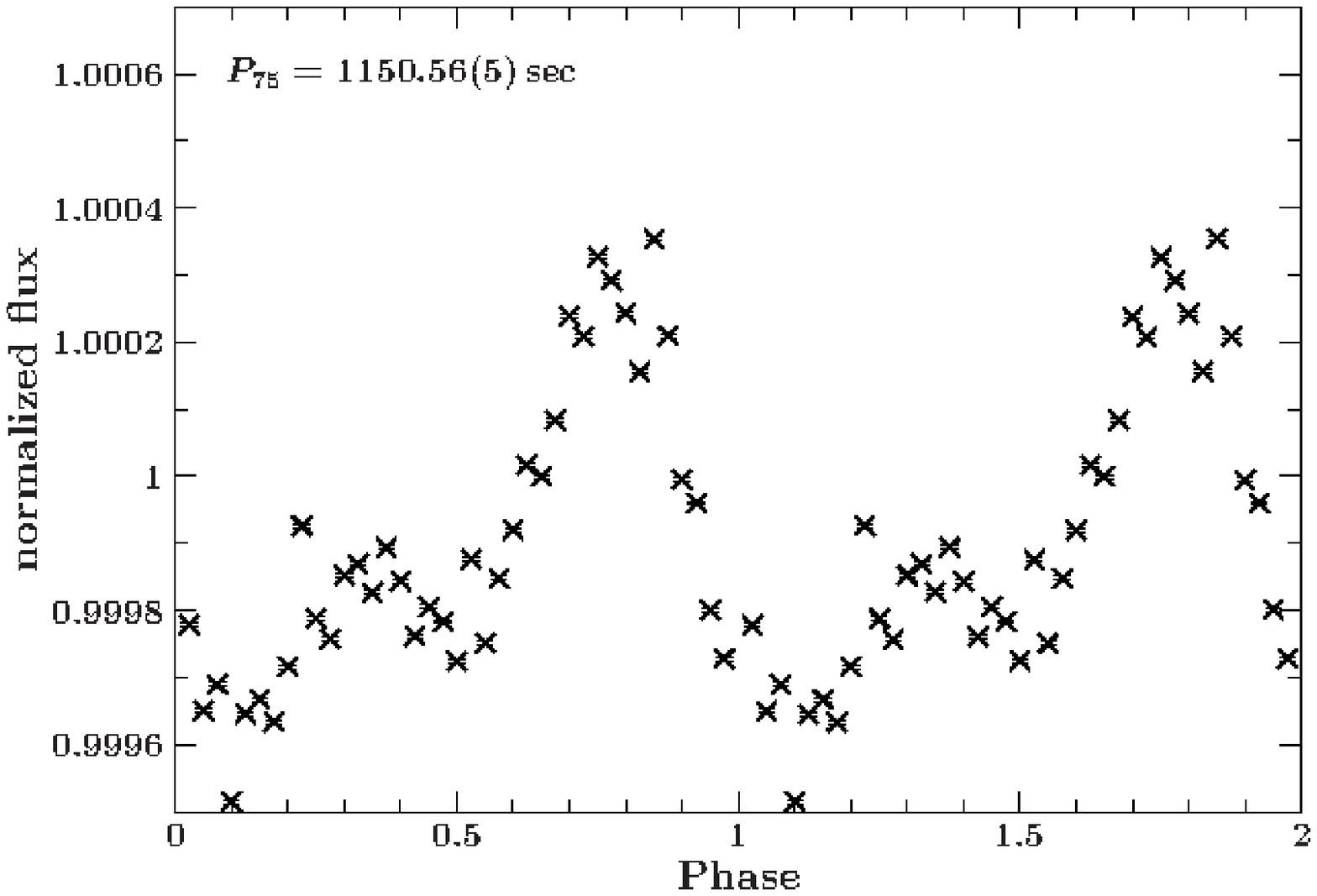}
\hspace*{0.05cm}
\includegraphics[width=0.485\textwidth]{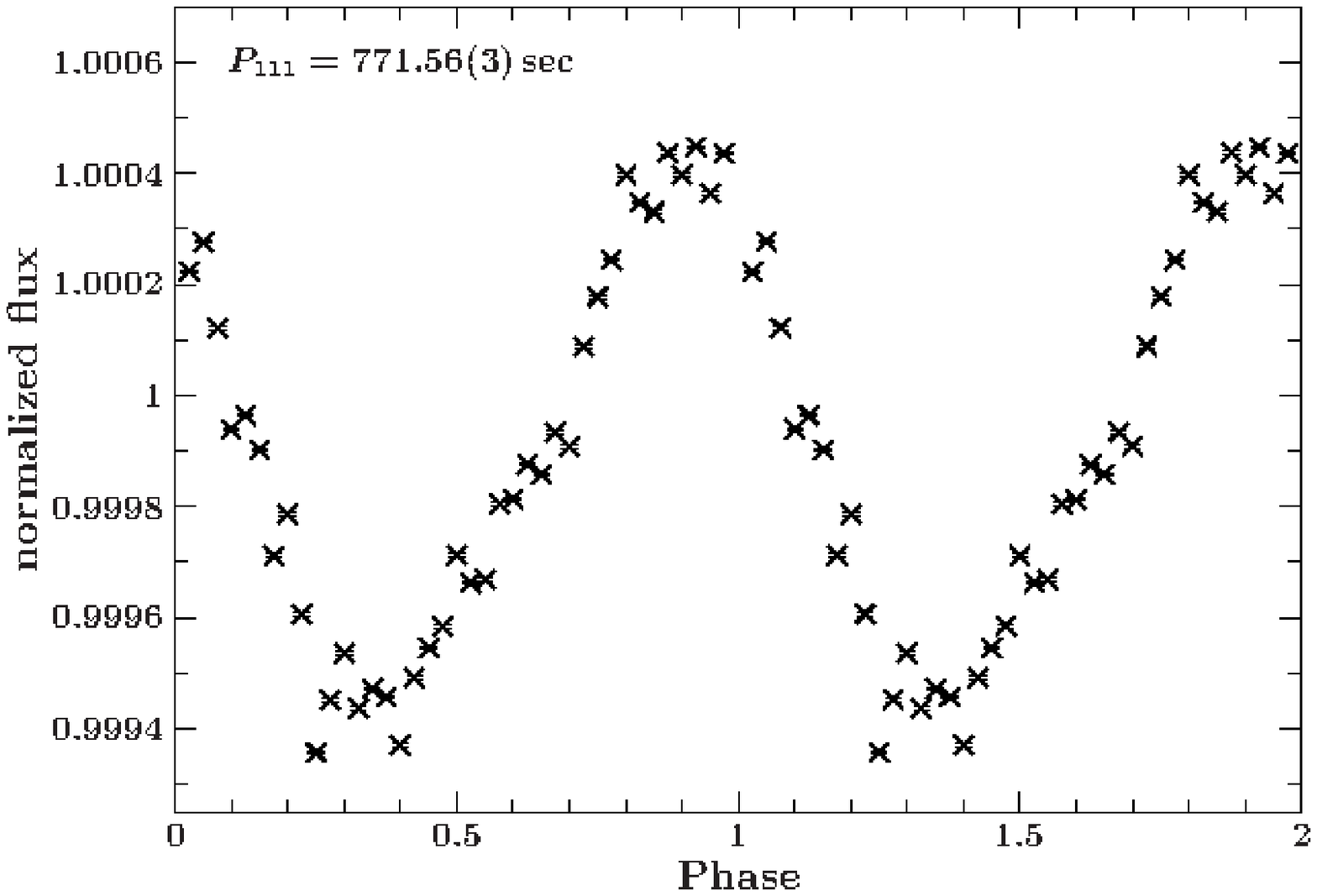}
\includegraphics[width=0.485\textwidth]{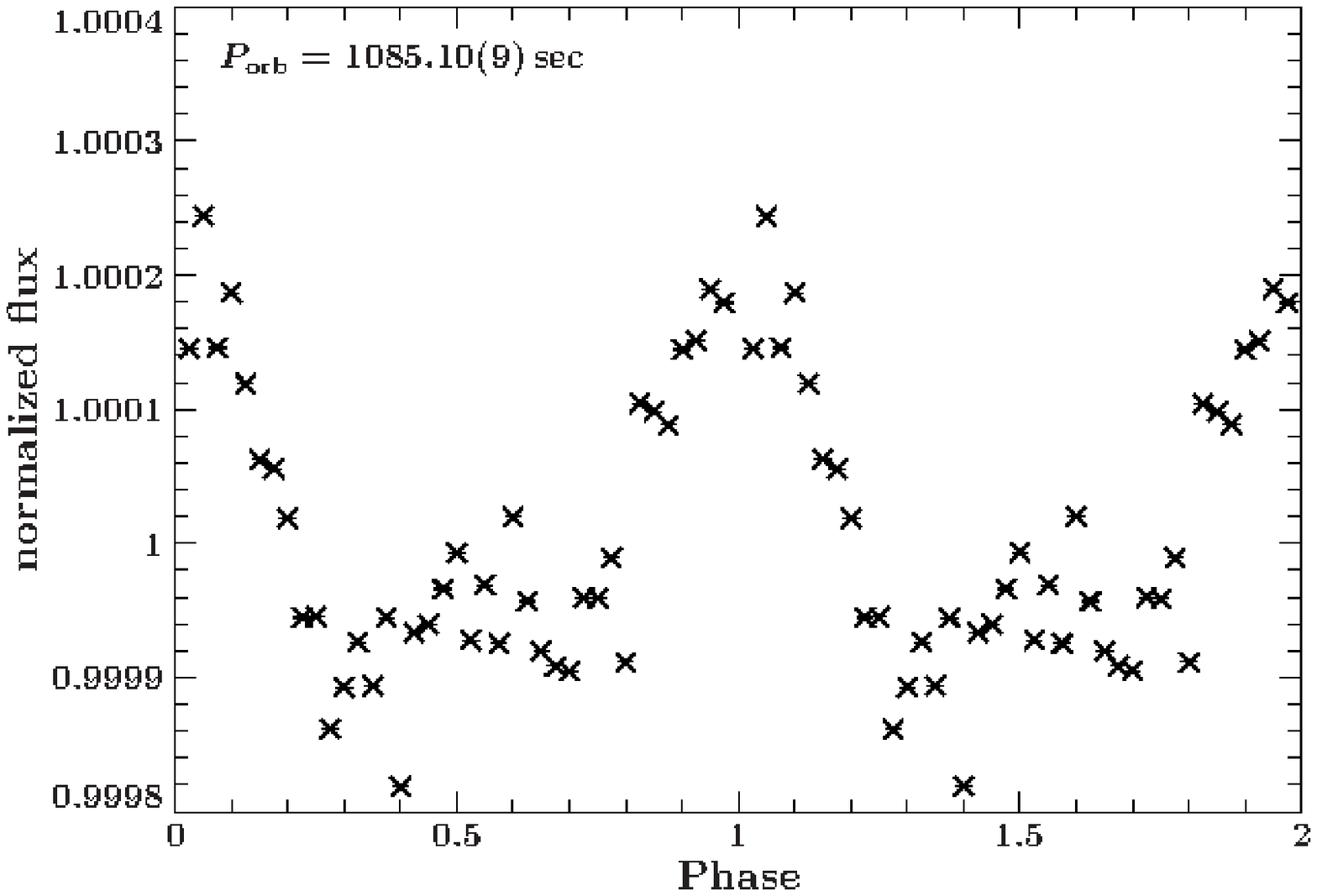}
\caption{Mean lightcurve of the five strongest variations detected in the lightcurve as well as the spectroscopic orbital period.}
\label{fig:mean}
\end{center}
\end{figure*}  

The frequencies around $90.63$ and $92.06$\,c/d show amplitude and frequency variations ($3$rd and $4$th panel in Fig.\,\ref{fig:fourier_maps}). The frequency variations of $90.63$ c/d moves opposite to the frequency variation of $92.06$\,c/d. The signal at $90.63$\,c/d becomes strongest at around BJD$-2454833=1200$\,days and shows an amplitude of $\sim700$\,ppm. The signal at $90.63$\,c/d becomes strongest at around BJD$-2454833=1000$\,days and reaches an amplitude of $\sim720$\,ppm.

The strongest frequency at around $184.13$\,c/d ($6$th panel in Fig.\,\ref{fig:fourier_maps}) shows a strong frequency and amplitude variability. After about one year the amplitude reaches its maximum with an amplitude of $\sim3350$\,ppm. After about another $200$ days the frequency splits up and almost disappears before it reaches a large amplitude again.

In particular in the panel with the orbital period ($2$nd panel in Fig.\,\ref{fig:fourier_maps}), many frequencies can be seen which show a similar amplitude but are only visible for a few weeks/months and therefore are below the detection threshold in the FT of the full lightcurve. For example, in the $2$nd panel of Fig.\,\ref{fig:fourier_maps} a peak starts to appear at about BJD$-2454833=1020$\,days at a frequency of around $79.66$\,c/d and stays for about 150\,days. This peak has a similar amplitude as the orbital period. However, because of the short duration this peak does not appear in the FT of the full lightcurve. Several more examples of frequencies which are only visible for a few weeks/months can be find over the full frequency range. Therefore, we note that the list of detected frequencies in Tab. \ref{tab:periods} has to be taken with care as most peaks are not stable and the given frequency can only be seen as average frequency over the observed period. Additionally, we provide the de-trended full {\it Kepler} lightcurve in electronic form with this paper. 

To refine the orbital period and derive the variability of the five strongest frequencies O-C diagrams were computed. The analysis of the O-C diagrams was done following the method outlined in Sec.~\ref{sec:ocanal}. The O-C diagram for the peak at $184$\,c/d was best reproduced with a period of $P_{\rm 184}=469.22(7)$\,sec. The O-C diagrams for the lower frequency peaks at $90$ and $92$\,c/d were best fitted with a period of $P_{\rm 92}=938.45(6)$\,sec and $P_{\rm 90}=953.29(0)$\,sec respectively. The two other frequencies detected at $74$ and $111$\,c/d were best reproduced with a period of $P_{\rm 111}=771.56(3)$\,sec and $P_{\rm 75}=1150.56(5)$\,sec. The orbital period can be best reproduced with a period of $P_{\rm orb}=1085.10(9)$\,sec. Table\,\ref{tab:refine} summarizes the periods measured from the O-C diagrams.

All six periods should be seen as average periods over the full observing period. Fig.\,\ref{fig:o-c} shows the derived O-C diagrams. In particular three periods ($P_{\rm 184}, P_{\rm 92}$ and $P_{\rm 90}$) show strong variations up to $\vert\dot{P}\vert\sim1.0\,$x$\,10^{-8}\,$s\,s$^{-1}$ over the observed timescale of 1052\,days. \citet{ski99} found a similar variation with similar strength in AM\,CVn itself in the superhump period. The other three periods ($P_{\rm 111}, P_{\rm 75}$ and $P_{\rm orb}$) show short term variations but no strong parabolic trend over time. 

\section{Mean phase-folded lightcurves of the periodic signals}\label{sec:mean_light}
The {\it Kepler} lightcurve was phase-folded on different periods and the mean phase-folded lightcurve was computed. Fig.\,\ref{fig:mean} shows the mean phase-folded lightcurves of the five strongest variations in the FT as well as the spectroscopic orbital period. The strongest can be seen at $P_{\rm 184}=469.22(7)$\,sec. The amplitude of this variation is about 5 times stronger than the second strongest variation at $P_{\rm 111}=771.56(3)$\,sec. The variations at $P_{\rm 90}=953.29(0)$\,sec and $P_{\rm 92}=938.45(6)$\,sec also show sinusoidal variability. The latter one is superimposed with the strong first harmonic $P_{\rm 184}$. The period at $P_{\rm 75}=1150.56(5)$\,sec shows two maxima whereas the first one is about twice as strong as the second one. The overall shape of the phase-folded lightcurve for $P_{\rm 75}$ looks similar to the superhump period found in HP\,Lib \citep{pat02}.

The mean phase-folded lightcurve folded on the spectroscopic orbital period ($P_{\rm orb}=1085.10(9)$\,sec; Fig.\,\ref{fig:mean}, lower left panel) shows a flat part followed by an increase and decrease in luminosity covering half of the orbit. \citet{lev11} presented a similar lightcurve for PTF1\,J071912.13+485834.0 ($P_{\rm orb}=1606.2\pm0.9$\,sec). They concluded that the increase in luminosity is most likely caused by the bright spot rotating in and out of the line of sight on the side of the accretion disc. We calculated the ephemeris where the zero phase is defined as when the lightcurve folded on the spectroscopic orbital period reaches its flux maximum:
\begin{equation}\label{equ:bjd}
BJD_{\rm max}=2455820.006744(2)+0.0125591(27)·E 
\end{equation}

\section{Discussion}
\subsection{Metal absorption lines}
Lines of various metals, like nitrogen, oxygen or carbon can be used to trace the evolutionary history of the system. Different abundance ratios link to different donor types. A high nitrogen to carbon/oxygen ratio is expected for a helium WD donor, whereas a helium star donor is expected to show higher carbon and oxygen abundances \citep{nel10}.

\citet{kup13} discovered strong absorption lines of magnesium and silicon in the three systems known to have orbital periods between $50$--$60$\,min. In the long period system GP\,Com, absorption lines of nitrogen were detected as well (Kupfer et al. in prep). So far, metal absorption lines have not been observed in short period systems. SDSS\,J1908 is the first high state system which shows a variety of metal lines, including N\,{\sc ii}, Si\,{\sc ii/iii} and S{\sc ii}, in absorption. Silicon as well as the sulphur lines can be used as tracer for the initial metalicity since their abundances are not supposed to be affected by nuclear synthesis processes during binary evolution. Remarkable is the detection of a large number of N\,{\sc ii} lines and the absence of oxygen and carbon lines. This might favour a helium white dwarf donor in SDSS\,J1908. However, this is only a qualitative statement, a detailed abundance analysis is necessary to prove the helium white dwarf nature of the donor star.

\subsection{Change of the orbital period}
\begin{figure}
\begin{center}
\includegraphics[width=0.48\textwidth]{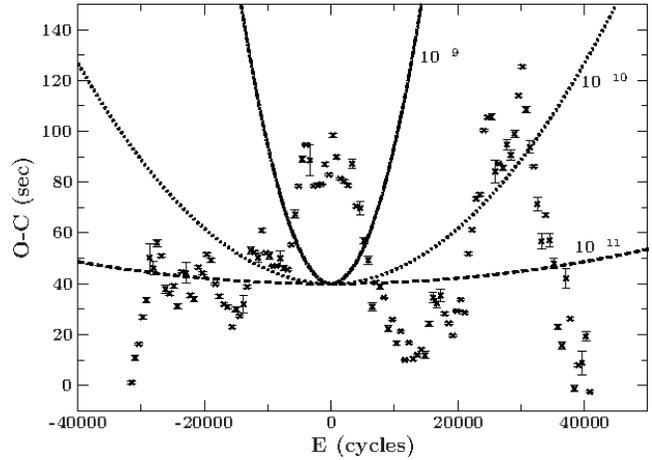}
\caption{O-C diagram of the orbital evolution in SDSS\,J1908. Overplotted are calculated O-C diagrams for different $\dot{P}$ (dashed line: $\dot{P}$=$10^{-11}$s\,s$^{-1}$, dotted line: $\dot{P}$=$10^{-10}$s\,s$^{-1}$, dashed-dotted line: $\dot{P}$=$10^{-09}$s\,s$^{-1}$)} 
\label{fig:pdot_orb}
\end{center}
\end{figure}

AM\,CVn type systems that passed the period minimum are expected to show an increasing period, $\dot{P}>0$ \citep{mar04}. In an O-C diagram an increasing period shows up as parabolic trend. We do not detect an overall parabolic trend in the O-C diagram of the orbital period (Fig.\,\ref{fig:pdot_orb}) but can set a rough limit on the $\dot{P}$. 

A formal fit to the data shown in Fig.\,\ref{fig:pdot_orb}, results in a $\dot{P}=9.6\times10^{-11}$s\,s$^{-1}$. For illustrative purposes we overplotted the expected signal for $\dot{P}$=$10^{-11}$-$10^{-09}$s\,s$^{-1}$. A conservative estimate rules out any orbital variations $>10^{-10}$s\,s$^{-1}$ which is well in agreement with the expected orbital increase of $\sim10^{-13}$s\,s$^{-1}$. 



\subsection{Origin of the photometric variations}
A Fourier analysis of the Q6 to Q17 short cadence data obtained by {\it Kepler} revealed a large number of frequencies above the noise level with most of them showing a large variability in frequency and amplitude. Some periods (e.g. $184.13$\,c/d) are visible over the full observing period, whereas many periods are only visible for a few weeks/months. 

In combination with phase resolved spectroscopy we were able to identify the orbital period $P_{\rm orb}=1085.10(9)$\,sec. The shape of the phase folded lightcurve at the period $P_{\rm 75}=1150.56(5)$\,sec looks similar to the superhump period found in HP\,Lib \citep{pat02}. Therefore $P_{\rm 75}$ corresponds most likely to the positive superhump period in SDSS\,J1908. This result leads to a period excess ($\epsilon=\frac{P_{\rm sh}-P_{\rm orb}}{P_{\rm orb}}$) of $\epsilon=0.0603(2)$, which is higher than observed in AM\,CVn ($\epsilon=0.0219$) and HP\,Lib ($\epsilon=0.0148$). \citet{pat05} found an empirical relation ($\epsilon=0.18q+0.29q^2$) between the period excess and the mass ratio ($q=\frac{M_2}{M_1}$) for a large number of hydrogen rich dwarf novae. Here, a mass ratio for SDSS\,J1908 of $q$=$0.33$ using this relation is obtained, which is much larger than found in AM\,CVn itself \citep{roe06a}. However, we note that the  empirical relation between the period excess and the mass ratio is possibly unreliable for AM\,CVn type systems \citep{pea07}. 

The shape of the phase folded lightcurve on the period $P_{\rm 92}=938.45(6)$\,sec and the O-C diagram looks like the variations found for the superhump period in AM\,CVn itself \citep{ski99}. Additionally, in AM\,CVn the strongest variation in a FT diagram corresponds to the first harmonic of the superhump period which is also the case for the $P_{\rm 92}$ period. If $P_{\rm 92}$ is the negative superhump period, this would lead to a very large period excess $\epsilon=0.1351(5)$ which was to our knowledge never observed in a system with a white dwarf accretor. Additional explanations for $P_{\rm 92}$ with its first harmonic could be the rotational period of the white dwarf. However, it is very unlikely that the rotational period of the accreting white dwarf shows strong period variations as found in the O-C diagrams over the course of 3 years. 

So far, we have only explained the origin of some of the strongest variations detected in the FT of SDSS\,J1908. The majority of observed frequencies in SDSS\,J1908 remains puzzling. Additional sources for variability which are visible over the observed period of about 3 years could be the rotational period of the accretor, pulsations in the accretor or variability in the disc. The first one can only explain a small number of additional periods detected in SDSS\,J1908 because there is only one rotational period of the accretor. The latter two could possibly explain a large number of periods as well as the appearance of frequencies which are only visible for a few weeks/months.

\citet{her14} found a large number of variations with pulsation periods between $828.2$ - $1220.84$\,s in the pulsating DAV white dwarf GD\,1212. Some pulsations in GD\,1212 show a large frequency variability similar to what we find for SDSS\,J1908. Pulsating DB white dwarfs (V777 Herculis stars) have effective temperatures between $21\,500$-$29\,000$\,K and show small amplitude variations and pulsation periods between $100$-$1100$\,sec \citep{cor12}. Although these systems show frequency-stable variability, all of our unexplained periods fall in that period range and therefore might be explainable with pulsations of the accretor which would be the first amongst the AM\,CVn systems. Alternatively, the variability of the frequencies and amplitude can be explained by variability in the disc of SDSS\,J1908. 


\begin{figure}
\begin{center}
\includegraphics[width=0.48\textwidth]{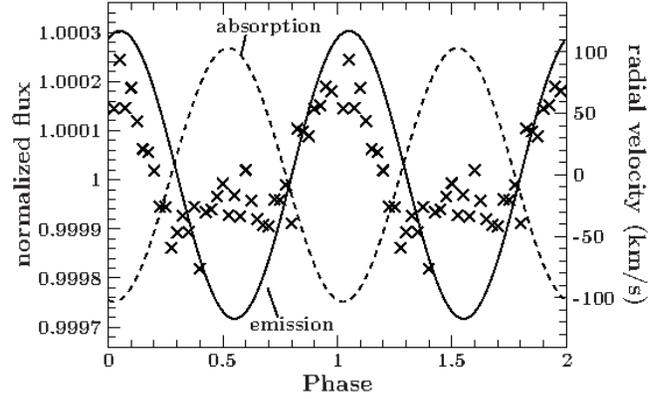}
\caption{Comparison between the phase folded lightcurve at the orbital period and the velocity variations of the He\,{\sc i} 4471\,\AA\,emission line (solid line) and the  average of the absorption lines (dotted lines) using the same ephemeris}    
\label{fig:comp_light_vel}
\end{center}
\end{figure}  

\subsection{Origin of the emission and absorption features and structure of the system}
We found two different variations in the spectroscopic data. Some of the helium lines (He\,{\sc i} 4713, 5047\,\AA), the Si\,{\sc ii} and the Mg\,{\sc ii} lines show a variation in absorption which moves with an offset of $170\pm15^\circ$ to the emission feature seen in the core of the He\,{\sc i} 4471\,\AA\,line. 


To test whether the absorption or emission feature are caused by motion of the bright spot, the donor or the accretor we compared absorption and emission lines to the variation in the lightcurve folded on the orbital period. 

In Fig.\,\ref{fig:comp_light_vel}, we present a binned, phase-folded photometric lightcurve together with the radial velocity curves of the emission from the He\,{\sc i} 4471\,\AA\,line and the absorption lines. The radial velocity curve and the phase-folded photometric lightcurve was folded on the ephemeris given in equation\,\ref{equ:bjd}. The maximum in brightness appears at the same phase when the absorption feature is most blueshifted. The maximum redshift of the He\,{\sc i} 4471 emission feature appears just after the brightness maximum. In the following, we discuss different scenarios to see if the phase offset between the lightcurve and the radial velocity curves can be explained. 


\textbf{1) Bright spot:} The variation in the lightcurve at the orbital period is caused by the bright spot. The maximum in brightness corresponds to the point when the bright spot is pointed to the observer. \citet{lev11} found for PTF1\,J071912.13+485834.0 a similar shaped photometric variation on the orbital period and showed that the maximum brightness happens when the radial velocity curve of the hot-spot crosses from blue- to redshifted, hence when the hot-spot is closest to us. Neither the emission feature nor the absorption lines are closest to us during maximum brightness and a similar explanation as for PTF1\,J071912.13+485834.0 fails. 




\textbf{2) Reflection effect:} The variation in the lightcurve at the orbital period is caused by a reflection effect of the irradiated side of the donor star. The hemisphere of a cool donor star facing the accretion disc is heated up by the significantly hotter accretion disc. This causes a variation in the lightcurve. More(less) flux is emitted if the irradiated hemisphere of the cool donor star is faced towards (away from) the observer. The maximum of brightness corresponds to the point when irradiated side of the donor star points towards to the observer.


\textbf{3) Gravity darkening:} The variation in the lightcurve at the orbital period is caused by gravity darkening on the highly deformed side of the donor star. The hemisphere of a cool donor star facing the accretion disc is highly deformed and the radius is much larger compared to the back side of the donor star which is not deformed. As a result, the back side of the donor star has a higher surface gravity, and thus higher temperature and brightness as well. The maximum of brightness corresponds to the point when back side of the donor star points towards to the observer. 

For all three scenarios we find no satisfying solution where any obvious feature (e.g. bright spot, donor star or accretor) is expected to be seen at the observed phases for the emission and absorption lines. 

Additionally, some helium lines (He\,{\sc i} 4387, 4921\,\AA), the N\,{\sc ii} and the Si\,{\sc iii} lines show no variation on the orbital period at all which means that they have to originate close the center of mass. In a typical AM\,CVn type system, the center of mass is close to the accreting white dwarf and the inner hotter parts of the accretion disc. Indeed the excitation energy for N\,{\sc ii} and Si\,{\sc iii} is similar ($\sim18$-$19$\,eV) and much higher than the excitation energy of Si\,{\sc ii} and the Mg\,{\sc ii} ($\sim8$-$10$\,eV). Therefore the origin of the N\,{\sc ii} and Si\,{\sc iii} lines could be in a distinct region in the inner hotter disc where the orbital motion is below our detection limit and the origin of the Si\,{\sc ii} and the Mg\,{\sc ii} could be in a distinct region in the cooler outer region of the disc where the orbital motion is higher.



\section{Conclusions and Summary}
The average spectrum shows strong helium absorption lines, typical for an AM\,CVn seen in high state. Additionally, a variety of weak metal lines of different species are detected. The phase-folded spectra and the Doppler tomograms reveal radial velocity variations at a period of $P_{\rm orb}=1085.7\pm2.8$\,sec which is in excellent agreement with a period at $P_{\rm orb}=1085.10(9)$\,sec detected in the three year {\it Kepler} lightcurve. Therefore, we identify $P_{\rm orb}=1085.10(9)$\,sec as the orbital period and prove the ultracompact nature of SDSS\,J1908.

A Fourier analysis of the Q6 to Q17 short cadence data obtained by the {\it Kepler} satellite revealed a large number of frequencies with strong variability in frequency and strength. In an O-C diagram we show that some periods show a strong variability similar to the superhump period of AM\,CVn itself. Some of the phase folded lightcurves of different periods show an overall shape similar to what is found for the superhump periods of HP\,Lib and AM\,CVn. Although some periods show very similar overall shape and variations in an O-C diagram compared to other high state systems, we are not able to identify unambiguously the negative or positive superhump in SDSS\,J1908.

The phase folded lightcurve on the spectroscopic orbital period shows a flat part followed by an increase and decrease in luminosity covering half of the orbit. \citet{lev11} found for PTF1\,J071912.13+485834.0 a similarly shaped photometric variation on the orbital period and showed that the variation is caused by the orbital period. However, in a comparison between the observed phases of the emission/absorption lines and the variation in the lightcurve we are not able to match the emission or absorption to any obvious feature in the binary such as the bright spot, the accretor or the donor star. Therefore the location of the spectroscopic variability remains undetermined.

\section*{Acknowledgments}
TK acknowledges support by the Netherlands Research School of Astronomy (NOVA). TRM and DS acknowledge the support from the Science and Technology Facilities Council (STFC) during the course of this work. PJG wishes to thank the California Institute of Technology its hospitality and support during a sabbatical leave.
\\
Based on observations made with the Gran Telescopio Canarias (GTC), installed in the Spanish Observatorio del Roque de los Muchachos of the Instituto de Astrof´ısica de
Canarias, in the island of La Palma. Based on observations with the William Herschel Telescope operated by the Isaac Newton Group at the Observatorio del Roque de los Muchachos of the Instituto de Astrofisica de Canarias on the island of La Palma, Spain. Some of the data presented herein were obtained at the W.M. Keck Observatory,
which is operated as a scientific partnership among the California Institute of Technology, the University of California and the National Aeronautics and Space Administration. The Observatory was made possible by the generous financial support of the W.M. Keck Foundation. This paper lincludes data collected by the {\it Kepler} mission. The authors gratefully acknowledge the {\it Kepler} team and all who have contributed to enabling the mission. The {\it Kepler} data presented in this paper were obtained from the Mikulski Archive for Space Telescopes (MAST). Funding for the {\it Kepler} Mission is provided by NASA’s Science Mission Directorate.
\bibliography{refs}{}
\bibliographystyle{mn2e}


\appendix

\section{}

\begin{table}
{\small \caption{Overview on the frequencies detected in the 3 year lightcurve of SDSS\,J1908} 
\label{tab:periods}
\begin{center}
\begin{tabular}{ll} 
\hline
F        & P      \\
 (1/d)   &(sec)    \\ 
\hline\hline 
\\[-3mm]
     70.719    & 1221.737  \\
     75.094    & 1150.566   \\
     75.927    & 1137.934   \\
     76.121    & 1135.037   \\
     79.623    & 1085.108  \\
     90.635    & 953.275    \\
     92.067    & 938.451    \\
     96.036    & 899.662   \\
    110.952    & 778.714    \\
    111.981    & 771.563   \\
    127.520    & 677.539    \\
    127.926    & 675.391    \\
    131.895    & 655.067    \\
    132.331    & 652.906    \\
    138.729    & 622.797   \\
    148.868    & 580.381   \\
    150.187    & 575.283   \\
    159.247    & 542.555   \\
    163.784    & 527.525   \\
    166.754    & 518.129    \\
    181.275    & 476.623   \\
    182.701    & 472.904    \\
    183.703    & 470.324    \\
    184.132    & 469.229    \\
    187.074    & 461.849   \\
    202.613    & 426.429    \\
    203.009    & 425.597   \\
    204.044    & 423.438    \\
    217.339    & 397.536   \\
    219.994    & 392.738    \\
    223.961    & 385.781   \\
    224.399    & 385.029    \\
    255.852    & 337.696    \\
    296.109    & 291.784    \\
    314.593    & 274.640    \\
    316.028    & 273.394    \\
    388.180    & 222.577    \\
    391.507    & 220.686    \\
    408.087    & 211.720    \\
    428.004    & 201.867    \\
    440.445    & 196.165    \\
    587.273    & 147.121    \\
\hline \\[-3mm]
\end{tabular}
\end{center}}
\end{table}

\begin{table}
 \centering
 \caption{Measured equivalent widths in (m\AA) and limits of disc emission and photospheric absorption lines}
  \begin{tabular}{cccc}
  \hline
   &  \multicolumn{1}{r}{EW (m\AA)} &  \multicolumn{1}{r}{EW (m\AA)} & \multicolumn{1}{r}{EW (m\AA)}  \\
  Line & \multicolumn{1}{r}{WHT} & \multicolumn{1}{r}{GTC} & \multicolumn{1}{r}{Keck}  \\
   \hline\hline 
  He\,{\sc i}  4009/4026       & $^a$             & 3387 $\pm$ 15  &   4373 $\pm$ 25  \\
  He\,{\sc i}  4120/4143/4168  & $^a$             & 3364 $\pm$ 18  &   4239 $\pm$ 24  \\
  \smallskip
  He\,{\sc i}  4388            & 2511 $\pm$ 26    & 3031 $\pm$ 16  &   1437 $\pm$ 18  \\
  He\,{\sc i}  4437/4471       &  \multirow{2}{*}{4615 $\pm$ 23}   & \multirow{2}{*}{4692 $\pm$ 24}  &  \multirow{2}{*}{2880 $\pm$ 24}   \\
   \smallskip
  Mg\,{\sc ii} 4481            &     &   &     \\
   Si\,{\sc iii} 4552           &  151 $\pm$ 14  & 89 $\pm$ 11 &  83 $\pm$ 8   \\
  Si\,{\sc iii} 4567           &  47 $\pm$ 8   & 34 $\pm$ 10 & 128 $\pm$ 14    \\
  Si\,{\sc iii} 4574           &  19 $\pm$ 7   & 19 $\pm$ 9 &   X   \\
  N\,{\sc ii} 4601             &  125 $\pm$ 14   & \multirow{2}{*}{128 $\pm$ 13}  &  87 $\pm$ 9   \\
  N\,{\sc ii} 4607             &  102 $\pm$ 12   &                                &  83 $\pm$ 10   \\
  N\,{\sc ii} 4613             &  83 $\pm$ 16   & 31 $\pm$ 9   &   39 $\pm$ 8  \\
  N\,{\sc ii} 4621             &  154 $\pm$ 18   & 98 $\pm$ 11  &  90 $\pm$ 11   \\
  N\,{\sc ii} 4630             &  118 $\pm$ 10   & 111 $\pm$ 12 &  146 $\pm$ 13    \\
  N\,{\sc ii} 4643             &  71 $\pm$ 9   & 65 $\pm$ 9   &   91 $\pm$ 14   \\
  He\,{\sc ii} 4685            &  --47 $\pm$ 13   & --21 $\pm$ 6  &  --19 $\pm$ 6   \\
  He\,{\sc i}  4713            &  1251 $\pm$ 24    & 932 $\pm$ 19  &   1012 $\pm$ 17  \\
  N\,{\sc ii} 4779             &   X  & 25 $\pm$ 5   &  57  $\pm$ 12   \\
  N\,{\sc ii} 4788             &   X  & 26 $\pm$ 6   &  56  $\pm$ 11   \\
  N\,{\sc ii} 4803             &   X  & 23 $\pm$ 5   &  92  $\pm$ 10   \\
  Si\,{\sc iii} 4828           &   X  & 31 $\pm$ 8   &  X   \\
  He\,{\sc ii} 4859            &   X  & 45 $\pm$ 10  &  $^b$   \\
    \smallskip
  He\,{\sc i}  4910/4921       &  3651 $\pm$ 26    & 2499 $\pm$ 21  &   2606 $\pm$ 24    \\
  N\,{\sc ii}  5001 - 5015     & \multirow{2}{*}{1083 $\pm$ 23}    & \multirow{2}{*}{1514 $\pm$ 24}  & \multirow{2}{*}{1182 $\pm$ 19} \\
  \smallskip
  He\,{\sc i}  5015            &     &  &    \\
  Si\,{\sc ii} 5041/5056       &  \multirow{2}{*}{836 $\pm$ 26}    & \multirow{2}{*}{1254 $\pm$ 25} &  \multirow{2}{*}{712 $\pm$ 22}  \\
   \smallskip
  He\,{\sc i}  5047            &     &   &   \\
  S\,{\sc ii}  5606            &  $^a$    & $^a$  & 38 $\pm$ 9    \\
  S\,{\sc ii}  5639/5640       &  $^a$    & $^a$  & 65 $\pm$ 12    \\
  S\,{\sc ii}  5647            &  $^a$    & $^a$  & 29 $\pm$ 8    \\
  N\,{\sc ii}  5666            &  $^a$    & $^a$  & 87 $\pm$ 13    \\
  N\,{\sc ii}  5676/5679       &  $^a$    & $^a$  & 92 $\pm$ 10     \\
  N\,{\sc ii}  5686            &  $^a$    & $^a$  & 73 $\pm$ 9    \\
  N\,{\sc ii}  5710            &  $^a$    & $^a$  & 71 $\pm$ 9    \\
  He\,{\sc i}  5875            &  $^a$    & $^a$  & $^c$   \\
  He\,{\sc i}  6678            &  $^a$    & $^a$  & 908 $\pm$ 24    \\
  He\,{\sc i}  7065            &  $^a$    & $^a$  & 804 $\pm$ 26    \\
  He\,{\sc i}  7281            &  $^a$    & $^a$  &  $^d$   \\
     \hline
\end{tabular}
\begin{flushleft}
Lines marked with an X indicate that this line is not detectable in the spectrum obtained\\
$^a$ Spectrum does not extend to this wavelength\\
$^b$ Line present but insufficient SNR to measure\\
$^c$ Blended with Na-D lines\\
$^d$ Line present but contaminated with atmosphere.
\end{flushleft}
\label{tab:equi}
\end{table}

\label{lastpage}

\end{document}